\documentclass[
 reprint,
 amsmath,amssymb,
 aps,
floatfix,
]{revtex4-2}

\usepackage{graphicx}
\usepackage{dcolumn}
\usepackage{bm}
\usepackage{upgreek}

\usepackage{graphicx}
\usepackage[colorlinks=true, allcolors=blue]{hyperref}
\usepackage{caption}
\usepackage{subcaption}
\usepackage[table]{xcolor}
\usepackage[normalem]{ulem}

\begin{document}


\title{Unveiling the charge distribution of a GaAs-based nanoelectronic device:\\ A large experimental data-set approach}
\thanks{Dataset for this article available at doi:10.5281/zenodo.6498343}%

\author{Eleni Chatzikyriakou\textsuperscript{1,}$^{\ddag}$, Junliang Wang\textsuperscript{2,}$^{\ddag}$, Lucas Mazzella\textsuperscript{2,3}, Antonio Lacerda-Santos\textsuperscript{1}, Maria Cecilia da Silva Figueira\textsuperscript{3}, Alex Trellakis\textsuperscript{3}, Stefan Birner\textsuperscript{3}, Thomas Grange\textsuperscript{4}, Christopher B\"auerle\textsuperscript{2}, Xavier Waintal\textsuperscript{1,}$^{\dagger}$}

\affiliation{\textsuperscript{1} PHELIQS, Université Grenoble Alpes, CEA, Grenoble INP, IRIG, Grenoble 38000, France\\ \textsuperscript{2} Univ. Grenoble Alpes, CNRS, Institut N\'eel, 38000 Grenoble, France\\ \textsuperscript{3} nextnano GmbH, Konrad-Zuse-Platz 8, 81829 M\"unchen, Germany\\ \textsuperscript{4} nextnano Lab, 12 chemin des prunelles, 38700 Corenc, France\\
$^\dagger$ corresponding author: \href{mailto:xavier.waintal@cea.fr}{xavier.waintal@cea.fr}\\
$^\ddag$ these authors contributed equally to this work
}

\date{\today}

\begin{abstract}
In quantum nanoelectronics, numerical simulations  
have become a ubiquitous tool. Yet the comparison with experiments is often done at a qualitative level or restricted to a single device with a handful of fitting parameters. In this work, we assess the predictive power of these simulations by comparing the results of a single model with a large experimental data set of 110 devices with 48 different geometries.  
The devices are quantum point contacts of various shapes and sizes made with electrostatic gates deposited on top of a high mobility GaAs/AlGaAs two-dimensional electron gas. We study the pinch-off voltages applied on the gates to deplete the two-dimensional electron gas in various spatial positions. We argue that the pinch-off voltages are a very robust signature of the charge distribution in the device. The large experimental data set allows us to critically review the modeling and arrive at a robust one-parameter model that can be calibrated in situ, a crucial step for making predictive simulations.
\end{abstract}

\maketitle

\section{Introduction}

As the field of quantum nanoeletronics becomes mature, the devices developed and techniques employed gain in complexity. The need for a set of predictive simulation tools is, therefore, becoming more acute. For instance, building a quantum computer requires a complete understanding of how single and multi-qubit properties depend on the geometry of the device as well as on the dynamical drives used to operate them. In the leading solid-state technology, superconducting based qubits, predictive simulation tools are already available. Very accurate models that involve only the electromagnetic degrees of freedom are already being used \cite{Koch2007, Krantz2019}. More importantly, the parameters of these models, \textit{i.e.} capacitances, inductances, and critical currents can be calculated or measured experimentally \textit{in situ}. The existence of such models has been critical in the development of superconducting based qubits. It allowed one to design several generations of quantum bits \cite{rasmussen2021}, to develop optimum strategies to drive them or entangle pairs of them, to explain quantitatively experimental data and understand the decoherence process. For semiconductor based quantum nanoelectronics, however, such predictive tools are not yet as advanced. With the work presented in this article, we aim at contributing to their development.

The difficulty in developing predictive simulation tools for semiconductor devices stems, at first, from the presence of very different length scales in the system. Indeed, the active quantum parts in such devices are typically much smaller than the length scales for which the electric potential is screened. It follows that there is a delicate interplay between the electrostatics of the system and the quantum mechanical response of the active part of the device. Consequently, these devices are much more sensitive to their microscopic environment \cite{Martinez2022,Percebois2021}. The large majority of quantum transport simulations simply ignore this difficulty. One assumes an effective form of the electric potential seen by the electrons and proceeds to calculate \textit{e.g.} the conductance of the system \cite{Bautze2014}. While this is sufficient to predict qualitative features, it suffers from severe limitations. First, the relation between the microscopic potential and the macroscopic parameters (gate voltages, sample geometry) is unknown. Therefore one needs to introduce various fitting parameters. Second, many effects, such as gate cross talk, are simply ignored. Third, the comparison to experimental data is (at best) limited to a single sample. Hence one cannot rely on such results to predict the behaviour of experimental devices. That is, the level of predictability of the simulations is difficult to assess. Finally, in some cases, \textit{e.g.} the quantum Hall regime at high magnetic field, the interplay between the electrostatic and the quantum problem leads to drastic reconstructions of the electrostatic landscape \cite{Chklovskii1992,Armagnat2020}. Self-consistent quantum-electrostatic calculations are therefore required even for qualitative results.

The need to improve the predictive power of the simulations has lead some groups to treat the electrostatic problem on the same level as the quantum one, see \textit{e.g.}  \cite{Wang2004,Marconcini2012,Woods2018,Armagnat2019,Woods2021}.  That is, of solving a set of equations capturing both the electrostatics of the system and the quantum behaviour of the active part of the device self-consistently. Such self-consistent simulations of quantum transport were for a long time confined to a few expert groups. Only recently commercial software such as nextnano \cite{nextnano2007} started to become available and open source codes were developed \cite{brandbyge2002,ferrer2014, groth2014,huang2016} in an effort to popularize the approach. More work remains, however, to be done on the modeling itself to improve and assess the predictive power of these simulations. Despite progress in the methods, the predictions can be very sensitive to the details of the modeling, such as the fraction of ionized donors, their distances to the active part of the device or the capacity of free surfaces to trap charges. 
With that many details potentially affecting the behaviour of the device, a match between the experimental data obtained from a single sample (often in a very narrow regime of gate voltages) and a numerical simulation of any transport properties is insufficient. It does not guarantee that one has properly captured the electrostatics of the device. Hence we argue that the weak point in current approaches is the feedback loop between experiments and simulations. That is, single (or few) sample studies are not enough to attain and demonstrate predictive power. The approach presented in this article relies on an extensive data set to put strong constraints on the modeling and assess its level of predictability. We implement this idea by designing specific experiments on well known systems for the sole purpose of validating the model used in the simulations. 

The experimental part of this work provides the extensive data set we use to calibrate the modeling and assess its predictive power. Indeed, as pointed out recently by Ref. \cite{Ahn2021}, there is a lack of extensive experimental measurements of nanoelectronic, quantum devices in the literature. Again, our objective is to assess how well we can predict {\it quantitatively} the behaviour of devices whose physics is supposed to be already well understood. We have fabricated a large set of quantum point contacts (QPC) on the two-dimensional electron gas (2DEG) formed in a GaAs/AlGaAs heterostructure. We have measured the low temperature differential conductance of a total of $110$ different quantum point contacts with $48$ different geometries of various shapes, widths and lengths. The full set of experimental data is published together with this article \cite{dataset}. Beyond the simulations presented here, such a database could be used in subsequent work as the modeling gets refined.

The simulation part predicts the different values of the gate voltages where the QPC conductance vanishes, the so-called ``pinch-off" voltages. Although in this article we restrain ourselves to predicting pinch-off voltages, our ultimate goal is to be able to perform parameter free simulations of electronic interferometers such as the one discussed in Ref. \cite{Bauerle2018, Edlbauer2022}. In fact, when modeling nanoelectronic devices such as the latter, an important aspect is the separation of energy scales in the system. In such samples, the relevant quantum physics takes place at energies of a few tens of $\upmu$eV. In contrast, the Fermi energy lies at much higher energies, \textit{i.e.} a few meV. At the same time, the macroscopic (gate voltages) or material (band offsets) parameters lie in the 1 eV range. Making predictive simulations is thus not straightforward. One must predict $\upmu$eV physics starting from a model that is only defined at much higher energies. The pinch-off values are unaffected by the low energy physics. Understanding them amounts to understanding the charge distribution in the device. That is, the physics in the meV--eV range. Only when one is confident that this physics is taken care correctly, it makes sense to try to predict the physics taking place at lower energies. That is, only when one can correctly predict the pinch-off voltage for any QPC among the $110$ devices fabricated to generate the data set, one can hope to develop a model precise enough such that it can correctly capture the relevant quantum physics. Hence, the simulations performed in this article aim at giving a quantitative answer to the question: ``where are the charges in the device?"

The article is organised the following way. Section \ref{sec:summary} summarizes our main findings: We show that the experimental pinch-off voltages match the predictions of the simulations within a $\pm 5\%$ accuracy. Section \ref{sec:exp} describes our experimental protocol. Section \ref{sec:sim} explains the model used in the simulations. In section \ref{sec:res} we present the comparison between the experimental data and the simulations. We end this article with section \ref{sec:discussion}, which contains a critical discussion of the modeling.

\section{Summary of the approach and main results}
\label{sec:summary}

We have fabricated and measured a large set of quantum point contacts of various shapes and sizes. Quantum point contacts are one of the simplest devices used in quantum nanoelectronics. Introduced in the seminal experiment demonstrating conductance quantization in a constriction \cite{Vanwees1988}, they became a standard tool to make tunable beam splitters \cite{Bauerle2018}. Despite their simplicity, there remain open questions about their behaviour in the regime called 0.7 anomaly \cite{Thomas1996}. Here, we do not focus on the 0.7 anomaly, nor on the quantization of conductance, but we rather establish, on firm grounds, the electrostatic potential seen by the conducting electrons. This amounts to understanding the charge distribution in the device. To reach this goal, we perform a systematic comparison between the simulated and measured ``pinch-off" voltages.

\begin{figure*}
    \centering
    \includegraphics[scale=0.55]{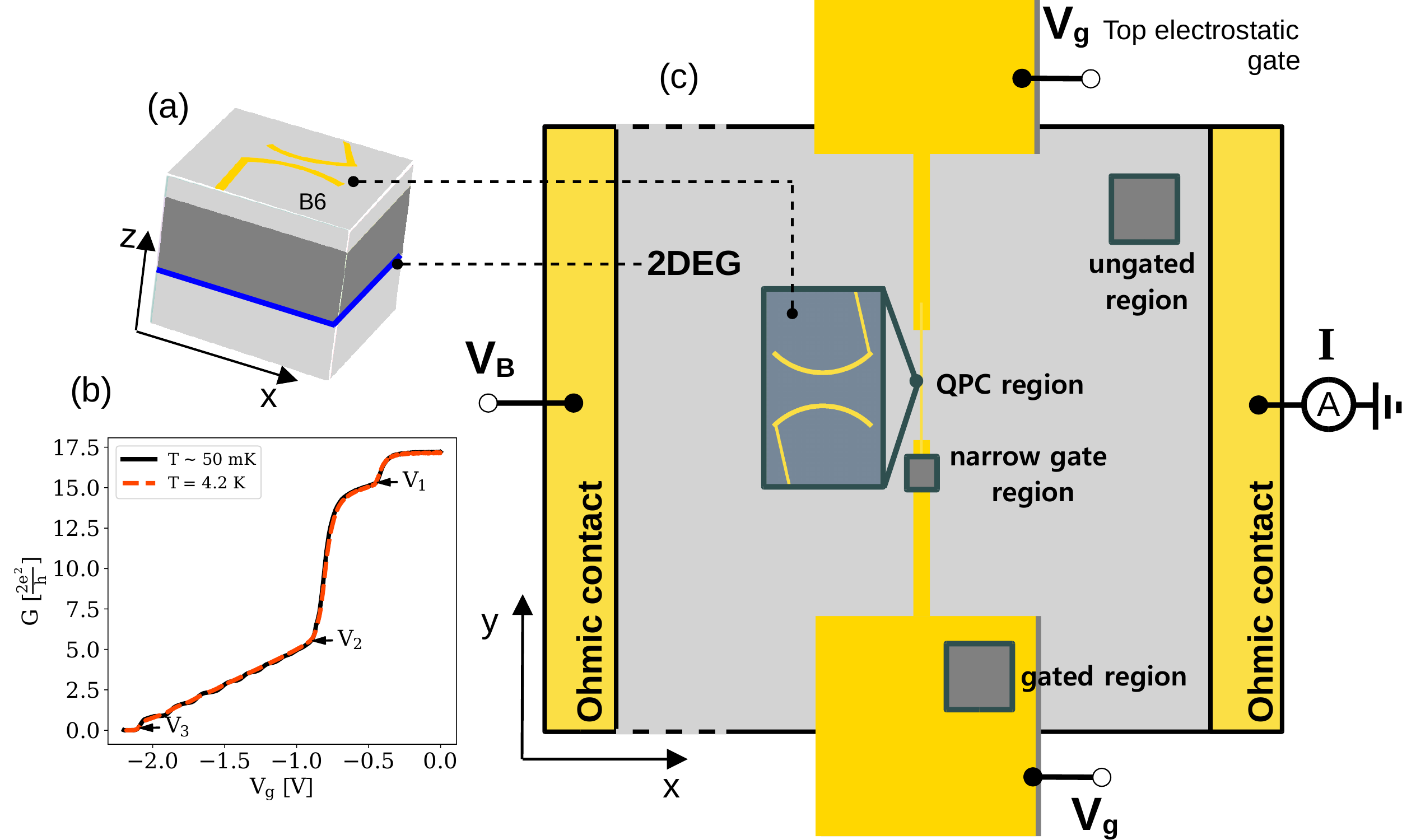}
    \caption{ 
    a) Schematic of the 3D stack with GaAs (dark gray), AlGaAs (light gray), top QPC gate (yellow) and the 2DEG region (blue). b) Typical experimental curve for gate voltage measurements. The three different points, $V_{\rm 1}$, $V_{\rm 2}$ and $V_{\rm  3}$, correspond to values of the gate voltage where the gas is depleted underneath the different gate regions. $V_{\rm  1}$ depletes the gas in the \emph{gated} region, $V_{\rm  2}$ in the \emph{narrow gate} region and $V_{\rm  3}$ in the \emph{QPC} region. 
c) Simplified top view of a device with a transistor-like geometry. 
    The Ohmic contacts (source and drain) and the electrostatic gates (situated $\approx$ 110 nm above the 2DEG) are indicated in yellow. 
    For the simulations, the system is broken into 4 different subregions tagged \emph{ungated}, \emph{gated}, \emph{narrow gate} and \emph{QPC} region, see text.}
    \label{fig:principle}
\end{figure*}

 Experimentally, the pinch-off voltage is the value of the voltage that needs to be applied to the electrostatic gates in order for the conductance to vanish or present a cusp --- an indication that the 2DEG gets fully depleted in some part of the system. 

Figure \ref{fig:principle}c shows a schematic of a typical device (see Fig. \ref{fig:SEM} for a SEM picture with the scales). The device (zoomed-in inset) has a transistor-like geometry with source and drain Ohmic contacts and electrostatic split gates. Applying a negative voltage $V_{\rm g}$ on the gates depletes the 2DEG underneath. As we indicate in Fig. \ref{fig:principle}c, each gate is further divided into three regions of different width. The region closest to the border of the 2DEG is very wide (several $\upmu$m) and is called the ``gated region".
 A second region of intermediary width (50 nm) is noted ``narrow gate'' region. Finally, the ``QPC'' region is located at the middle of the device where the gates split. A sketch of the full stack, a standard high mobility GaAs/AlGaAs heterostructure, is shown in Fig. \ref{fig:principle}a. 
 
 \begin{figure*}
    \centering
    \includegraphics[scale=0.55]{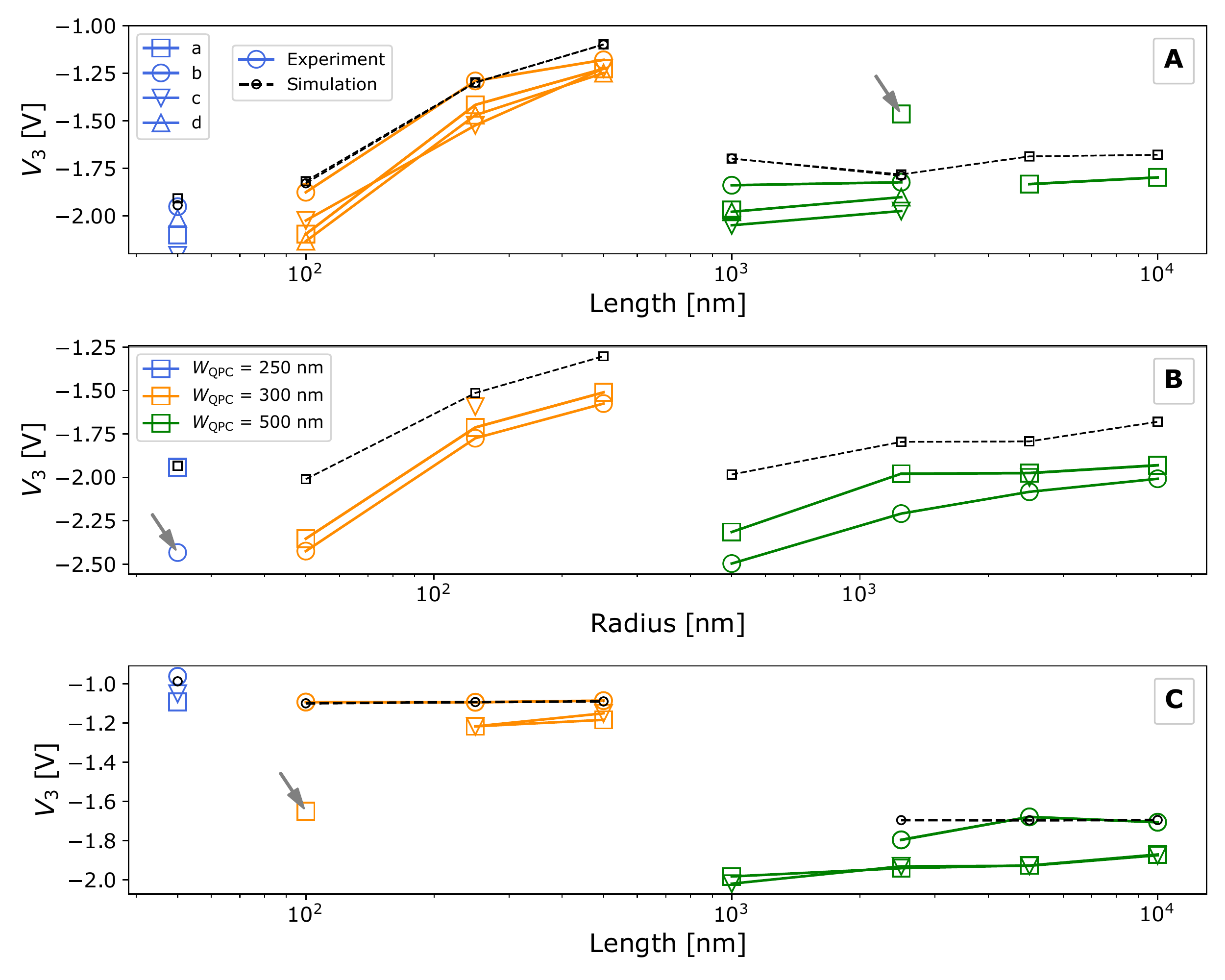}
    \caption{Comparison between simulation (small black symbols and dashed line) and experiment (color symbols and solid line). The QPCs are grouped according to design: A (top), B (middle), C (bottom). The results have been color-coded according to the width of the QPC, W\textsubscript{QPC} (Figure \ref{fig:Designs}): blue for 250 nm, orange for 300 nm and green for 500 nm. The different symbols correspond to different devices with identical nominal characteristics but at different locations in the wafer, c.f. Figure \ref{fig:sample_layout}. A letter is attributed to each symbol: ``a" (rectangles), ``b" (circles), ``c" (up triangles) and ``d" (down triangles) so that a given QPC is uniquely identified by its geometry (A, B, C or equivalently upper, middle and down panel), its rank (1-8 from left to right in the figure) and the letter a,b,c,d. For instance QPC ``A5b" corresponds to the fifth circle in the upper panel. Arrows point to outliers that we attribute to lithography problems or structural damage during cooldown or initial measurements, see text.}
    \label{fig:A1B1_A1C1_sim_exp}
\end{figure*}
 
 We measure the current I versus gate voltage $V_{\rm g}$ characteristic for each device; see Fig. \ref{fig:principle}b for a typical experimental trace. As one decreases $V_{\rm g}$ from zero towards negative values, one first depletes the 2DEG underneath the ``gated" region. Indeed, the large width of the gates (several $\upmu$m) on this region compared to the rest of the split gate means the 2DEG will first be depleted underneath it. The value for which the 2DEG is depleted underneath the ``gated'' region is denoted $V_{\rm 1}$. There, one observes a cusp in the current--gate voltage curve, as indicated on Fig. \ref{fig:principle}b. 

In the simplest model for $V_{\rm 1}$, accurate within a few percent (see the discussion in section \ref{sec:discussion}), the 2DEG and the electrostatic gate form a simple plane capacitor. The electron density in the gated region is given by $n(V_\mathrm{g}) = n_\mathrm{g} - \epsilon V_\mathrm{g}/(ed)$ ($n_\mathrm{g}$: electronic density in the gated region with zero volts applied to the gate, $\epsilon\approx 12\epsilon_\mathrm{0}$: dielectric constant, d=110 nm: total distance between the 2DEG and the gate). It follows that $V_\mathrm{1}$ is an almost direct measure of the  electronic density in the gated region,

\begin{equation}
 n_\mathrm{g} \approx \frac{\epsilon V_\mathrm{1}}{ed}.
\end{equation}

As one further decreases the gate voltage, one eventually depletes the gas below the ``narrow gate" region. This region is tens of micron long along the $y$ direction, but only 50 nm wide. 
A second cusp in the conductance versus $\rm V_g$ curve is observed at the voltage $V_{\rm 2}$ where this region is fully depleted. Finally, as one continues to decrease $V_{\rm g}$ towards strongly negative values, the gas is depleted in the central QPC region. At that moment the conductance between the left and right Ohmic contact vanishes entirely. We denote the gate voltage at which this  depletion is observed as $\rm V_3$. The set of voltages $V_{\rm 1}$, $V_{\rm 2}$ and $V_{\rm 3}$ reflect the initial density at various parts of the sample and the interplay between the field effect of the gate and the screening of the 2DEG. This is the main data studied in this paper. 
The full set of current-gate voltage characteristics is provided as a zenodo archive \cite{dataset}. They could be further used to study, \textit{e.g.} conductance quantization.

In order to predict the different values $V_{\rm 1}$, $V_{\rm 2}$ and $ V_{\rm 3}$, we perform a different type of calculation for each of the three gate regions. The dimensions of the ``gated'' and ``narrow gate'' regions have been kept constant for all QPCs. Hence we expect very little sample to sample variation of the experimental value for $\rm V_1$ and $V_{\rm 2}$. The value of $V_{\rm 3}$, however, corresponds to the ``QPC'' region that has been varied in different devices. 

\begin{itemize}

\item To calculate $V_{\rm 1}$, one simulates the ``gated" region. It can be approximated as infinite along $x$ and $y$ directions due to the large dimensions of the gates. Therefore one only needs to perform 1D simulations along the $z$ direction. Additional 1D simulations were performed for the ``ungated'' region, \textit{i.e.} without top gate. It allows one to calculate the 2DEG bulk density $n_{\rm s}$ far away from the gates. Such value can be compared to the experimental bulk density $n_{\rm bulk} \rm = 2.8\cdot 10^{15}$ $\rm m^{-2}$ obtained by Hall measurements. 

\item To calculate the value of $V_{\rm 2}$, we simulate the narrow gate region. The latter is very long along the $y$ direction (up to 50 $\upmu$m), but very narrow (50 nm in most samples) along $x$. Hence we consider a system infinite along $y$ and need only to perform 2D simulations in the ($x$, $z$) plane. We decrease $V_{\rm g}$ until the density vanishes underneath the middle of the \emph{narrow} gate. Then we record the associated value of $V_{\rm g}$ as $V_{\rm 2}$.  
\item To calculate the value $V_{\rm 3}$ we perform a full 3D simulation of the ``QPC'' region. The $\rm V_3$ value is then extracted by decreasing $V_{\rm g}$ until the density vanishes underneath the middle of the gap between the two gates. At $V_{\rm g} \le V_{\rm 3}$, the 2DEG is split into two disconnected left and right parts.

\end{itemize}

The model we used to simulate the devices has two {\it a priori} independent input parameters: the dopant density $n_{\rm d}$ and surface charges density $n_{\rm sc}$ (see Sec.\ref{sec:sim}). With this model we do not attempt to predict the experimental bulk 2DEG density in the ungated ($n_{\rm bulk}$) or gated ($\propto V_{\rm 1}$) regions. Instead, we calibrate the model values of $n_{\rm d}$ and $n_{\rm sc}$ by fitting the model to the experimental values of $V_{\rm 1}$ and $n_{\rm  bulk}$. This calibration sets the value of the electronic density in the model in the ungated ($n_{\rm s}$) and in the gated ($n_{\rm g}$) regions. While $n_{\rm s} = n_{\rm bulk}$ after calibration, we keep two different letters for the model and experimental values, respectively, for clarity. Predicting $n_{\rm s}$ and $n_{\rm g}$ would imply having a precise knowledge of many microscopic parameters. Accurate values of the dopant ionization energy, dopant concentration, surface states energy, band alignment, dielectric layers thickness etc. would have to be obtained either from theoretical arguments or from experiments. This is a hard task, and also not necessary for the physics we seek to understand, the transport properties. At the end of this article, we should argue that $n_{\rm d}$ and $n_{\rm sc}$ are in fact {\it not} independent and that a single effective input parameter may be used (see the discussion of Fermi level pinning in Sec. \ref{sec:discussion}). This further increases the predictive power of our model. However, the relation between $n_{\rm d}$ and $n_{\rm sc}$ has not been assumed in the simulations and is considered here as a prediction of the modeling.

Once $n_{\rm d}$ and $n_{\rm sc}$ have been calibrated, we then proceed to predict the $V_{\rm 2}$ and $V_{\rm 3}$ pinch-off voltages. The main result of this article is shown in Fig. \ref{fig:A1B1_A1C1_sim_exp} where we compare the experimental (color symbols) to the simulated (black symbols) values of $V_{\rm 3}$ for the different QPC designs; see Fig.\ref{fig:Designs} for the latter. It shows a systematic agreement of the theoretical prediction for $V_\mathrm{3}$ with that obtained experimentally within a precision of 10\% or better.  Figure \ref{fig:A1B1_A1C1_sim_exp} implies that we can reliably predict the spatial variations of the electronic density in devices of arbitrary geometries.  This opens the path for making quantitative calculations at smaller energy scales and predict genuine quantum effects quantitatively and without fitting parameters.
 
Beyond the overall agreement between experiments and simulations, Fig. \ref{fig:A1B1_A1C1_sim_exp} further shows significant sample to sample variations for nominally identical samples as well as systematic deviations (the simulation curves being systematically above the experimental ones). These features, which we attribute to disorder, will be discussed later in this article. 

\section{Experiments: details of the set of quantum point contact devices}
\label{sec:exp}
\begin{figure}
\centering
\includegraphics[width=0.49\textwidth]{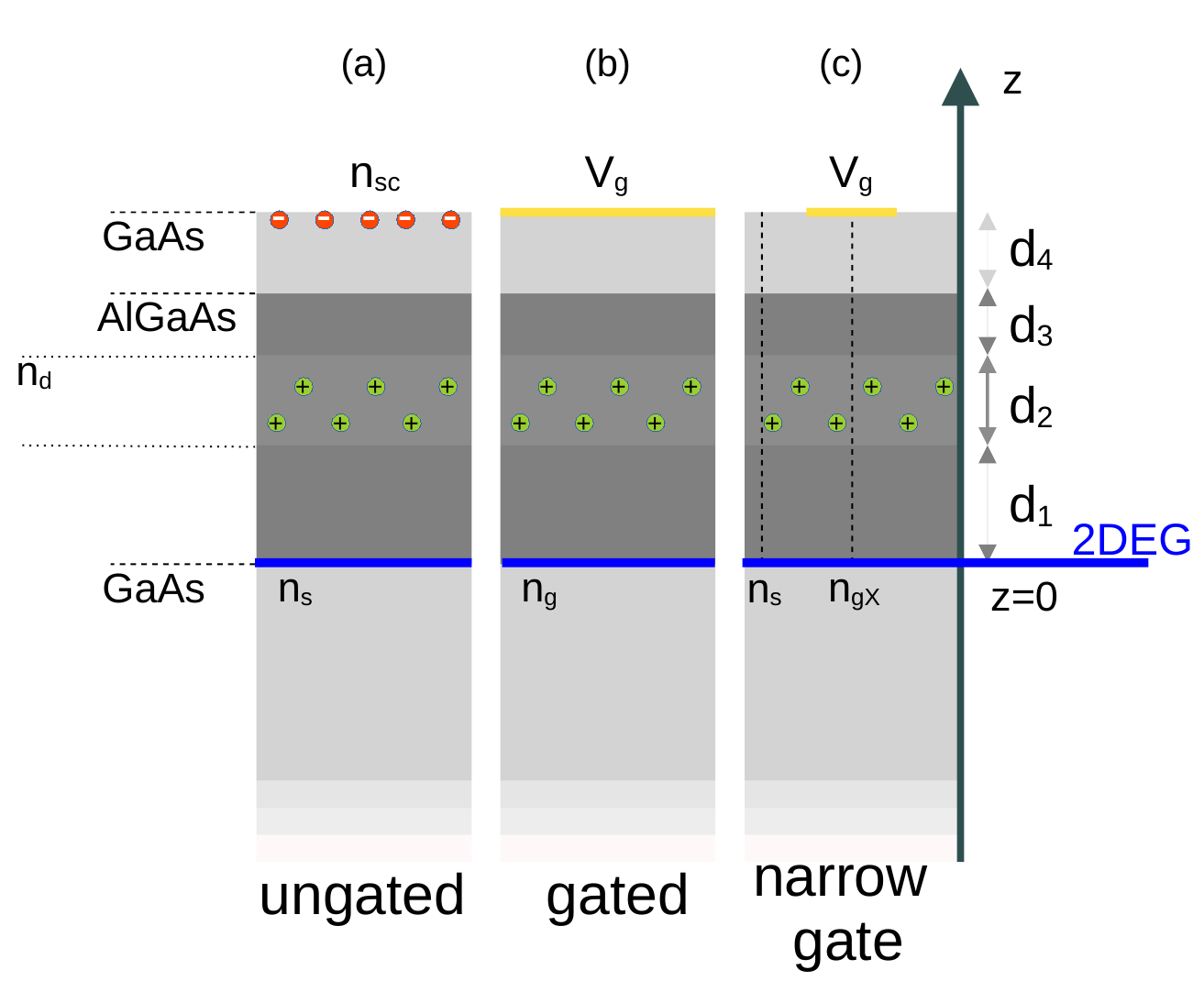}
\caption{Panel (a): Side view of the experimental heterostructure stack. The widths of the different layers are respectively $\rm d_1 = 25$ nm, $\rm d_2 = 65$ nm, $\rm d_3 = 10$ nm and 
$\rm d_4 = 10$ nm. The central AlGaAs layer of width $\rm d_2$ is doped. Panel (a), (b) and (c) correspond, respectively, to the ungated, gated and narrow gate regions as indicated in Fig. \ref{fig:principle}c.  In the simulations, (a) and (b) correspond to 1D models without and with a top gate, respectively, while (c) corresponds to a 2D model with a gate of finite width (50 nm) at its surface.}
\label{fig:HEMTs}
\end{figure}

Our samples were fabricated on a Si-modulation-doped GaAs/$\rm Al_{0.34}Ga_{0.66}$As heterostructure grown by molecular beam epitaxy (MBE). The high mobility two-dimensional electron gas (2DEG) lies at the GaAs/AlGaAs interface, located 110 nm below the surface. Performing Hall measurements at 4.2 K under dark conditions, we find a bulk 2DEG density of $n_{\rm bulk}\rm \approx 2.79 \times 10^{15}$ $\rm m^{-2}$  and a mobility of $\rm \mu \approx 9.1 \times 10^{5}$ $\rm cm^{2}$/Vs. The corresponding Fermi wave-length is $\lambda_{\rm F} = \sqrt{2\pi/n_\mathrm{s}} \approx$ 47 nm. The surface electrodes that define the quantum point contacts are made out of a metal stack of 4 nm titanium and 13 nm gold, deposited by successive thin-film evaporation. The composition of the stack of the heterostructure is shown in Fig. \ref{fig:HEMTs}a together with the widths of the different layers.

\begin{figure}[!]
\centering
\includegraphics[width=0.45\textwidth]{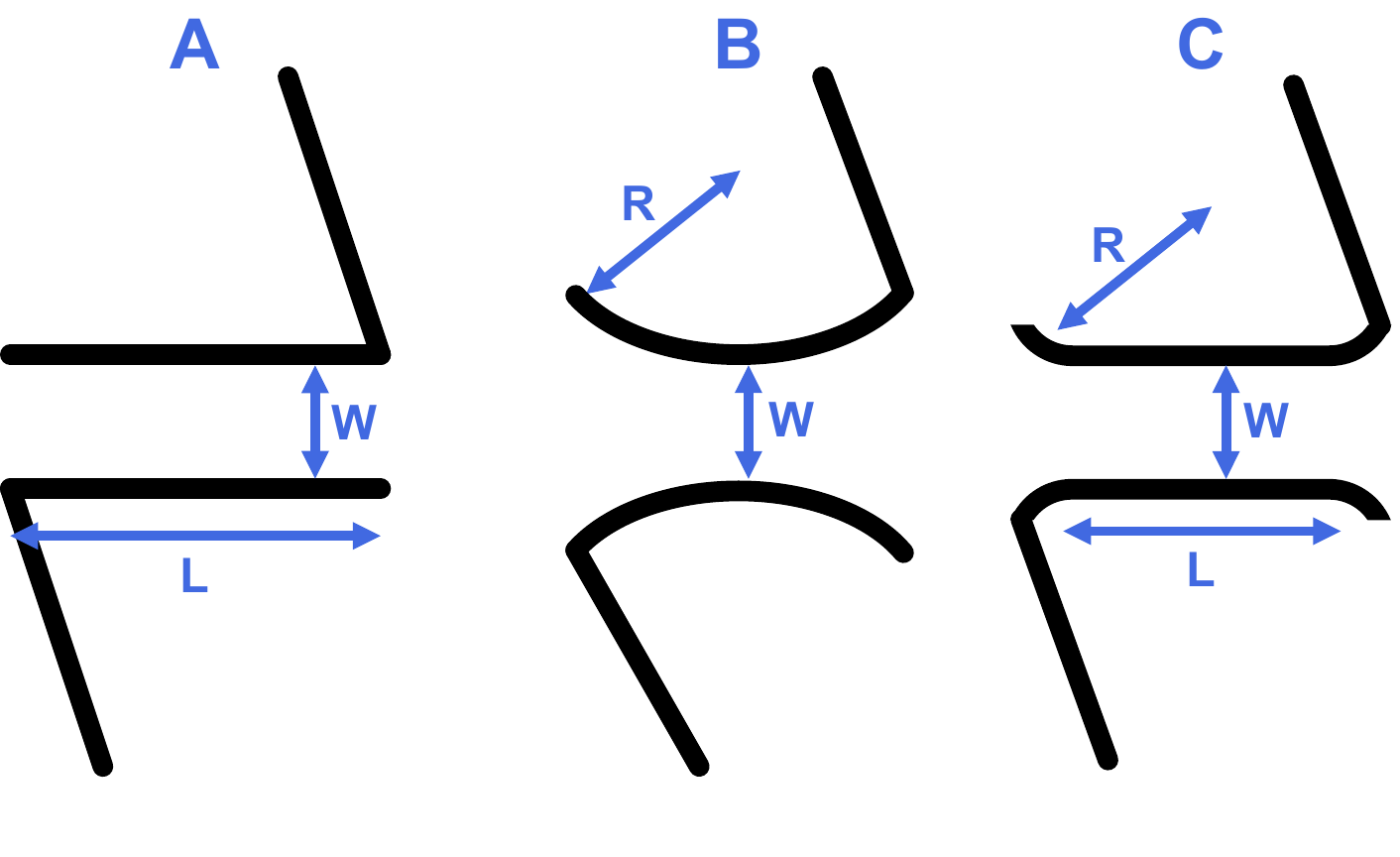}
\caption{Schematic of QPC designs: Rectangular (A), Round (B) and Smooth (C). The characteristic geometrical parameters $L$ (length), $W$ (width) and $R$ (radius) are indicated by arrows.}
\label{fig:Designs}
\end{figure}

In order to investigate the geometrical influence of QPCs, we designed three kinds of shapes: Rectangular (A), Round (B) and Smooth (C) (see Fig. \ref{fig:Designs}). 
Rectangular (A) designs correspond to a wire of length $L$ defined by two parallel gates separated by width $W$. Round (B) designs consist on two semi-circular gates with radius $R$ that define the point contact. At last, Smooth (C) designs belong to an intermediate design between A and B, combining the linear constriction with adiabatic entrances.

For each design (A,B,C), 16 different combinations of geometrical parameters $L$, $R$ and $W$ are investigated, from the smallest (A1, B1, C1) to the largest (A16, B16, C16) sizes. 

Figure \ref{fig:SEM} shows Scanning Electron Microscopy (SEM) images of various fabricated designs; see Appendix \ref{sec:suppl} for exact parameters.
To account for statistical variability, devices with the exact same design are repeated across the chip. We label them with an additional Latin letter (``a" to ``d") in the device name. For example, A2a and A2b are different QPCs with identical nominal characteristics. 

\begin{figure*}
    \centering
    \includegraphics[scale=0.8]{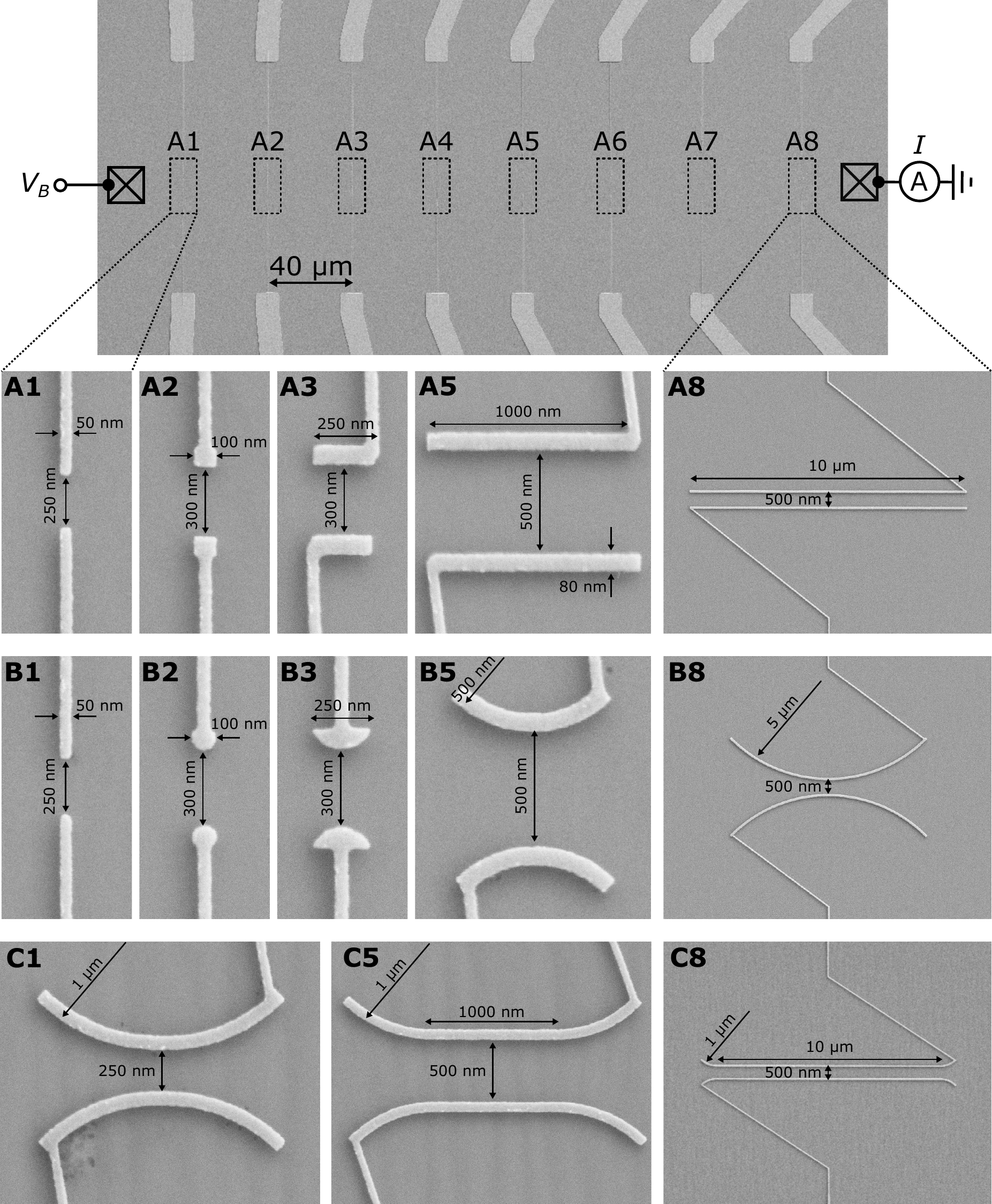}
    \caption{Scanning electron microscopy (SEM) images of quantum point contacts. (Top) Overview of a set of 8 QPCs in series sharing a pair of Ohmic contacts (a sample). (Bottom) Examples of investigated shapes (A, B, C).}
    \label{fig:SEM}
\end{figure*}

\begin{figure}
    \centering
    \includegraphics[scale=0.49]{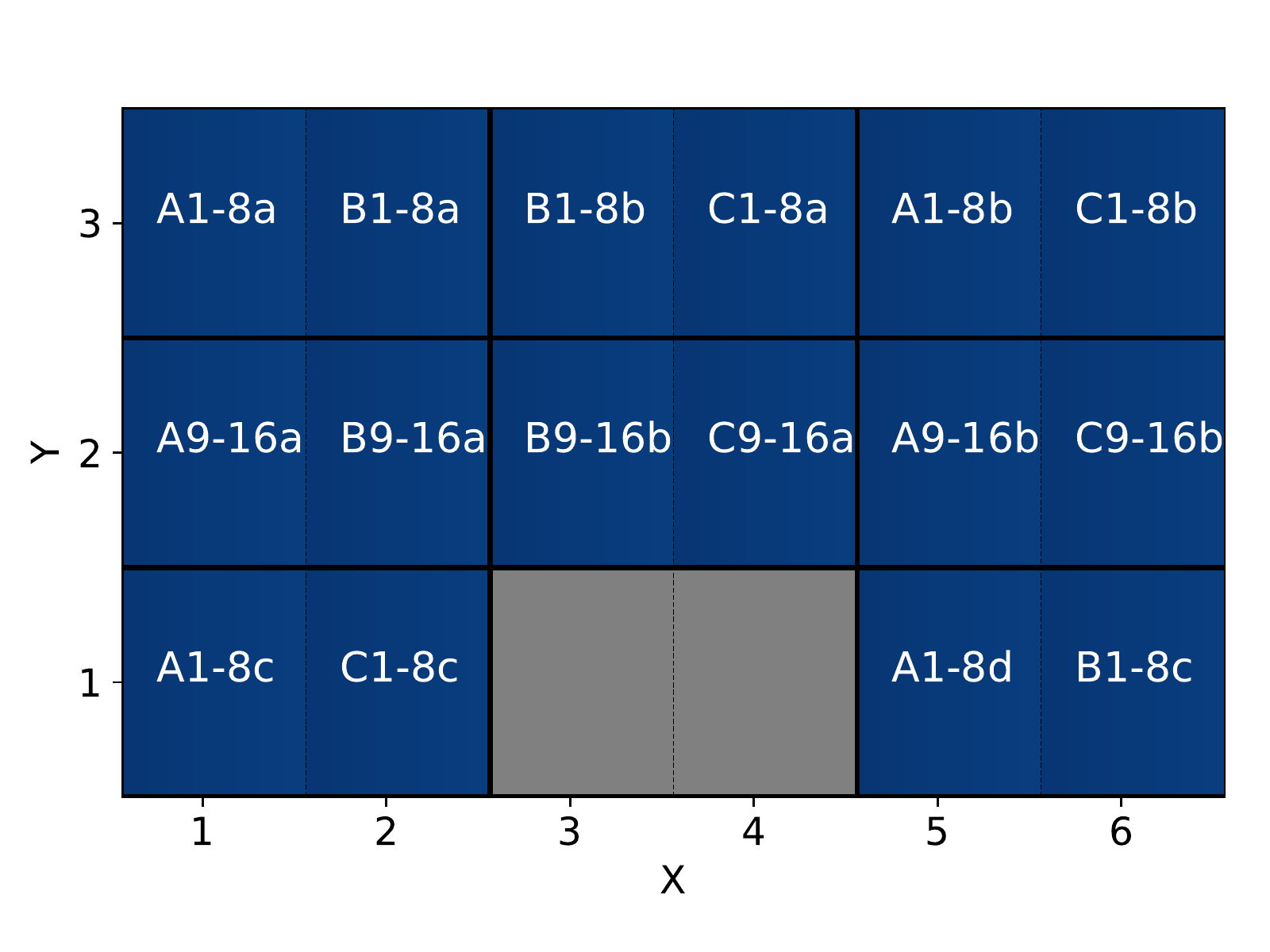}
    \caption{Repartition of the QPCs in the GaAs dice. A set of 8 QPCs (a sample) is represented as a rectangular box identified by its X and Y indices. Gray areas have not been used. The dimensions of one sample are $\rm \sim 1.6\times 2.3$ $\rm mm^2$.}
    \label{fig:sample_layout}
\end{figure}

\begin{figure*}
\centering
\includegraphics[width=0.95\textwidth]{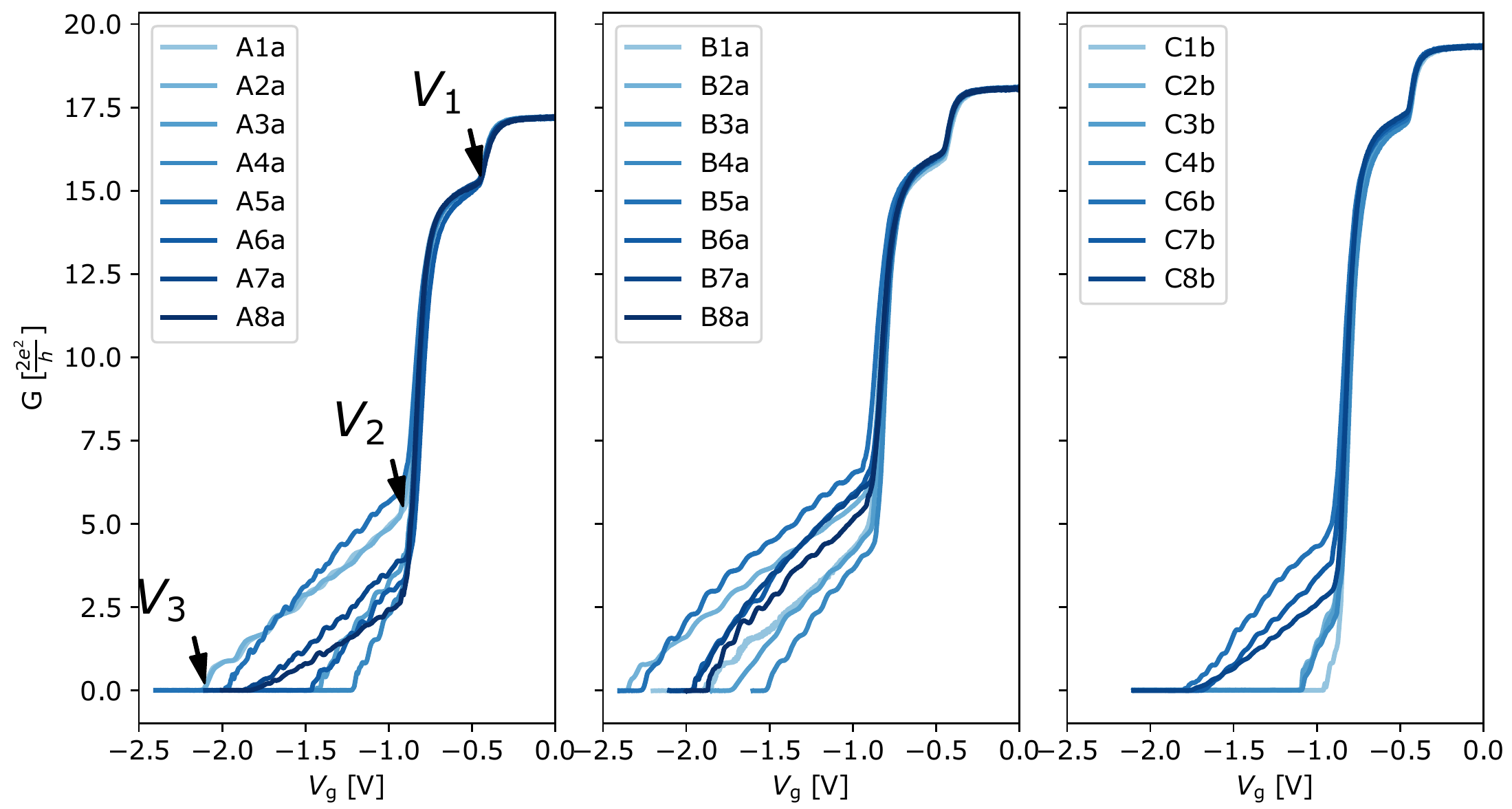}
\caption{
G = I/$V_\mathrm{B}$ versus $V_\mathrm{g}$ measurements for 3 sets of QPCs (A1-8a, B1-8a, C1-8b) with an Ohmic bias $V_\mathrm{B}\rm = 500 \upmu V$ at T $\rm \approx$ 50 mK. The arrows indicate the characteristic voltage drops $V_\mathrm{1}$, $V_\mathrm{2}$ and $V_\mathrm{3}$. Note that the current $I_\mathrm{0}(V_\mathrm{g}=0)$ is the same for all QPCs in a given set. This is because of the common contribution from the Ohmic contact.}
\label{fig:IV_exp}
\end{figure*}

In order to maximize the number of measured devices in a same cooldown, a set of 8 QPCs is placed in series sharing a common pair of Ohmic contacts (see top panel in Fig. \ref{fig:SEM}). With a separation more than 40 $\upmu$m, we ensure that no mutual effect occurs between the neighbouring QPCs. We call such a set of 8 QPCs, a sample. We draw attention to the fact that we follow this notation throughout the text, as different such sets of QPCs (samples) present larger deviation in their measured characteristics than QPCs within the same sample.

We fabricated and measured a total of 110 QPCs with 48 unique designs that are distributed in 16 sets on a chip of 10 mm $\times$ 8 mm. A schematic layout is shown in Figure \ref{fig:sample_layout}. The sample that contains a given QPC can be identified by the a column index X and a row index Y. 
For example, the device A2a is located in the set X=1 and Y=2.

The conductance characterization was performed at two temperatures T $\approx$ 4.2 K and $\rm T\approx$ 50 mK. Unless stated explicitly, all the data shown below have been taken at 4.2 K as only a limited number of samples have been measured at 50 mK. While the temperature strongly affects features like conductance quantization, the temperature variations of the pinch-off voltages can be ignored, as one can observe in Fig. \ref{fig:principle}b. We note, however, that there is a small decrease of $\rm \le$ 25 mV of the $V_{\rm 3}$ pinch-off voltages between 4.2 K and 50 mK. This small variation is irrelevant here considering the level of accuracy of the simulations and the sample to sample experimental variations. We apply a bias voltage $V_{\rm B}=500  \, \upmu$V between the Ohmic contact to induce the current I. To characterize the transport properties, we measured the current I as a function of surface-gate voltage $V_{\rm g}$ for each device. The full data set of these transport measurements, can be found in \cite{dataset}.

Figure \ref{fig:IV_exp} shows conductance versus $V_{\rm g}$ measurements for various QPCs at 50 mK temperatures, which have more pronounced quantization features than those at 4.2\,K. Three distinct regions can be identified separated by the pinch-off voltages $V_{\rm 1}$, $V_{\rm 2}$ and $V_{\rm 3}$. In the first two regions ($V_{\rm g} \ge V_{\rm 2} \approx$ -0.75 V), different devices share the same conductance behaviour. This is expected as in this regime the current is dominated by the electron flow in the large ``gated" or the ``narrow gate" regions, which is identical for all QPCs (see Fig. \ref{fig:principle}). 
In the third region ($V_{\rm g} \le V_{\rm 2}$), the transport properties are only affected by the narrow constriction formed between the gates. Clear conductance quantization steps are observed for numerous QPCs with wide-ranging pinch-off voltages $V_{\rm 3}$. Note that the pinch-off voltages $V_{\rm 1}$ and $V_{\rm 2}$ are also visible when one biases only one of the two gates (e.g. top or bottom).
Also note that we show the raw data without substraction of the series resistance due to the Ohmic contacts and measuring apparatus.

A few samples deviated significantly from the theoretical predictions, as indicated by the grey arrows in Fig. \ref{fig:A1B1_A1C1_sim_exp}. We have performed a visual inspection of the SEM image of some of these samples which did not reveal any particular problem. We attribute these outliers to fluctuations of the density in the QPC region due to \textit{e.g.} a fluctuation of the concentration of dopants above.

This article focuses on the proper level of modelisation to capture spatial variations of the electronic density. We leave to future work the analysis of more subtle features such as the shapes and positions of conductance plateaus. As a general trend, we find, in accordance with common knowledge, that the plateaus get quickly washed out upon increasing the temperature to 4.2 K or making the sample too long (only the  ones with $L\rm \le 250$ nm
showed clear plateaus). Type C samples showed the least pronounced plateaus features.

\section{Simulations: details of the modeling}
\label{sec:sim}

\begin{figure*}
   \centering
     \begin{subfigure}{0.95\textwidth}
         \centering
         \includegraphics[width=\textwidth]{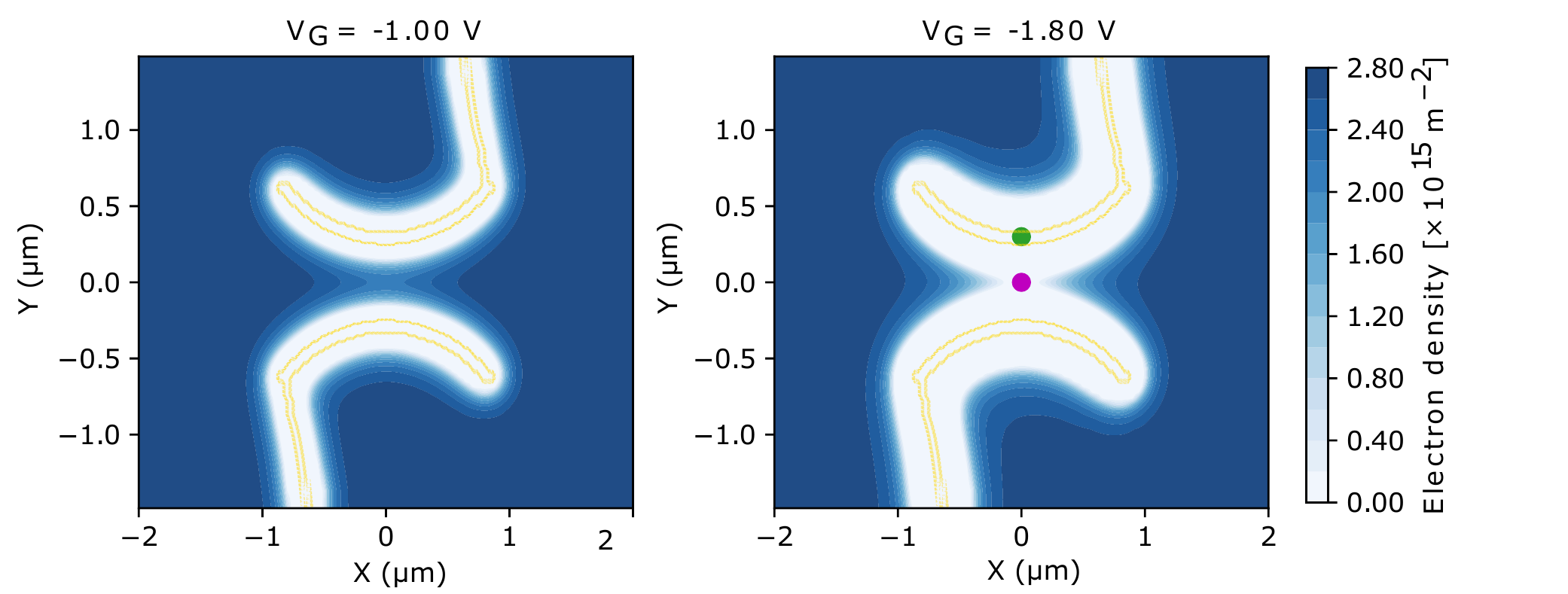}
         \caption{}
         \label{fig:2D_cut_B1_6}
     \end{subfigure}
    \centering
    \hfill
     \begin{subfigure}{0.9\textwidth}
         \centering
         \includegraphics[width=\textwidth]{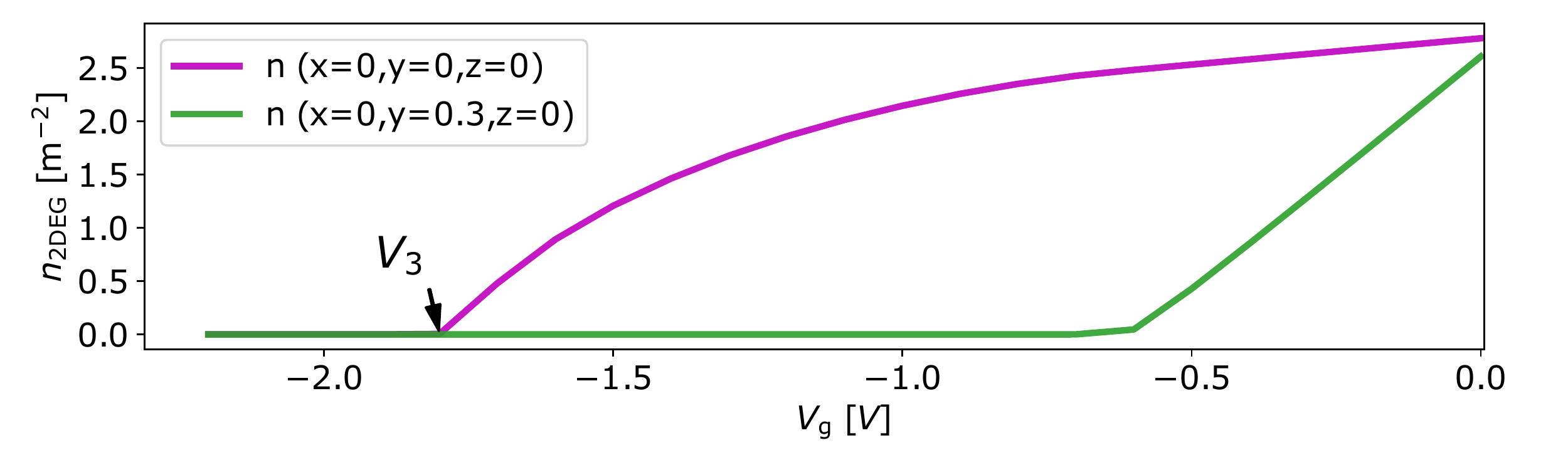}
         \caption{}
         \label{fig:B1_6_IV}
     \end{subfigure}
     \caption{(a) Simulation of the electron density distribution in the 2DEG for QPC B6 at different voltages. Left: $V_\mathrm{g}=-1.0$ V. Right: $V_\mathrm{g}$ = -1.8 V. (b) Density versus $V_\mathrm{g}$ at the two different points indicated in the right (a) panel. $V_\mathrm{3}$ is identified as the value for which n$(x=0,\ y=0,\ V_\mathrm{g} = V_\mathrm{3})$ vanished.}
\end{figure*}

The simulations performed in this article are done within the Thomas-Fermi approximation at zero temperature using the commercial software nextnano++ \cite{nextnano2007, Trellakis2007}. We model the device using the self-consistent Poisson equation,

\begin{equation} \label{eq:TF_equation_sec_4}
\vec\nabla . \left[ \epsilon(\vec r) \vec\nabla U(\vec r) \right] = e N[\mu = E_\mathrm{F}+eU(\vec r) ] - e N_{\rm d}(\vec r)
+ e N_{\rm sc}(\vec r)
\end{equation}

where $U(\vec{r})$ is the electrostatic potential, $\epsilon(\vec{r})$ the dielectric constant, $N_{\rm d}(\vec{r})$ is the ionized dopant density in the doped layer, $E_\mathrm{F}$ = 0 the Fermi level (electro-chemical potential) of the 2DEG and $N_{\rm sc}(\vec{r})$ the frozen surface charge density at the surfaces not covered by metallic gates and $\mu$ is the chemical potential of the 2DEG. We use the uppercase letter $N$ to indicate volume densities (\textit{e.g.} density of dopants $N_{\rm d}$) in $\rm m^{-3}$ and the lowercase letter n to indicate surface densities in $\rm m^{-2}$ (\textit{e.g.} $n_{\rm s}$ the bulk 2DEG electronic density). Whenever possible, we will convert the volume densities to effective surface densities. For instance the dopant density $N_{\rm d}$ over a layer of thickness $\rm d_2$ is equivalent to an effective 2D dopant density of $n_{\rm d} = N_{\rm d} d_2$. We use Dirichlet conditions $U(\vec{r}) = V_\mathrm{g} - V_\mathrm{w}$ at the gate-semiconductor interface. $V_\mathrm{g}$ is the applied voltage with respect to the grounded 2DEG. $V_\mathrm{w}$ is the work function of the gold/GaAs interface which for definiteness we take as $V_\mathrm{w}\ \approx$ 0.75 V. However, the actual value of $V_\mathrm{w}$ is actually irrelevant since any change of $V_\mathrm{w}$ will be compensated by a change in $n_\mathrm{d}$ to keep $V_\mathrm{1}$ calibrated to the experiments.

To complete the theoretical model we must provide the relation between the density N of the 2DEG and the chemical potential $\mu$.

 This relation is defined by the integral up to $\mu$ of the system local density of states which in general must be calculated by solving the quantum problem self-consistently with the Poisson equation \cite{Armagnat2019}. Here, we approximate the local density of states to be equal to the bulk density of states of GaAs, ignoring the quantum fluctuations (Thomas-Fermi approximation). The integrated DOS equation for $N$ thus reads: 
\begin{eqnarray}\label{eq:dosdos}
    N(\mu) &=& \frac{(2m^*)^{3/2}}{3 \pi^2\hbar^3} (\mu - E_\mathrm{b})^{3/2}   \  {\rm for } \  \mu > E_\mathrm{b} \nonumber \\
    N(\mu) &=& 0  \  {\rm for } \  \mu \le E_\mathrm{b}
    \end{eqnarray}
where $E_\mathrm{b}$ is the position of the bottom of the conduction band in GaAs and $m^*$ its effective mass. As discussed for $V_\mathrm{w}$ above, the actual value of $E_\mathrm{b}$ is irrelevant in this article. Note that for the purpose of pinch-off voltage calculations, we could have used the constant density of state of a 2DEG, $n = m^*\mu/(\pi\hbar^2)$, instead of the three dimensional Eq.\eqref{eq:dosdos} and obtained the same results within our accuracy.

Figures \ref{fig:HEMTs}a and \ref{fig:HEMTs}b show a side view of the geometry used in the simulations of the ``ungated" and ``gated" QPC regions respectively (see Fig. \ref{fig:principle}c). We define: $n_\mathrm{s}$ as the 2DEG density underneath the ungated region, and  $n_\mathrm{g}$ the 2DEG density underneath the gated region for $V_\mathrm{g}$ = 0. The stack is made of several layers of widths $\rm d_i$. The models for the gated and ungated regions are translationally invariant along the ($x$, $y$) plane, hence the problem reduces to a 1D simulation along the $z$ direction. Figure \ref{fig:HEMTs}c shows a side view of the geometry of the ``narrow gate" region. Since it is invariant only along $y$, the problem reduces to a 2D simulation of the ($x$, $z$) plane. Finally, Fig.\ref{fig:principle}a shows the geometry used for a QPC region. The simulations of the QPC regions are performed in 3D. A single set of parameters $n_\mathrm{d}$, $n_{\rm sc}$ and the thicknesses $\rm d_1= 25$ nm, $\rm d_2=65$ nm, $\rm d_3=10$ nm and $\rm d_4=10$ nm is used in the simulations of all the different regions.

The values of $n_\mathrm{s}$ and $n_\mathrm{g}$ at $V_\mathrm{g}$ = 0 is a complex function of the model parameters $n_\mathrm{d}$, $n_{\rm sc}$, $V_\mathrm{w}$, $E_\mathrm{b}$ and the $\rm d_i$. However, once these parameters are set (in our case, calibrated to the experiments), the density profile and the electric potential in the 2DEG are simply a function of $n_\mathrm{s}$, $n_\mathrm{g}$, $V_\mathrm{g}$ and the total distance $\rm d = \sum_{i=1}^4 d_i = 110$ nm between the 2DEG and the gates. The point of view taken in this article is to use $n_\mathrm{s}$ and $n_\mathrm{g}$ as effective parameters and ignore the large set of microscopic parameter. We do not attempt to describe the detailed microscopic physics that would allow one to predict their values. Note that in a typical 2DEG, $n_\mathrm{d}$ is roughly equal to 10 times $\rm n_s$, \textit{i.e.} 90\% of the dopant electrons go to the top surface and only 10\% to the 2DEG \cite{Buks1994b}. Furthermore, not all dopants necessarily get ionized. Hence a precise calculation of $n_\mathrm{s}$ (idem for $n_\mathrm{g}$) requires a very precise knowledge of the dopant density and of the various  energies level of the dopants and at the surface. 

In the simulations, we used a mesh with a discretization step smaller than 1 nm. We explicitly checked that the results are unaffected by the discretization within a precision better than 10 mV by performing several simulations with higher accuracy.

Figure \ref{fig:2D_cut_B1_6} shows a typical 3D simulation of a QPC region (here device B6) at different gate voltages. The color map shows the electronic density around the central part of the device. At $V_\mathrm{g} \gg V_\mathrm{3}$ , the density is only slightly decreased below the gates. 
At $V_\mathrm{g}\ =\ -1.8 V < V_\mathrm{3}$ , the region in between the two gates is fully depleted. Figure \ref{fig:B1_6_IV}, shows the density versus $V_\mathrm{g}$ at two different points of interest. As expected, we find that the pinch-off, \textit{i.e.} cutting the system into disconnected left and right parts, occurs when the central point $x$ = $y$ = 0 is depleted. Hence we take the corresponding $V_\mathrm{g}$ value as our calculated $V_\mathrm{3}$. The typical potential profile observed in the simulations is almost flat in the 2DEG and abruptly rises in regions where the 2DEG has been depleted and cannot screen the gates. Plots of the behaviour of the potential (at zero field but also in the quantum Hall regime) can be found in \cite{Armagnat2019}.  

\section{Comparison between Experimental and Simulation Pinch-off Voltages}
\label{sec:res}

\subsection{Model Calibration using the $\rm V_1$ pinch-off of the gated regions}
\label{sec:calV1}

\begin{figure}
    \centering
    \includegraphics[width=0.42\textwidth]{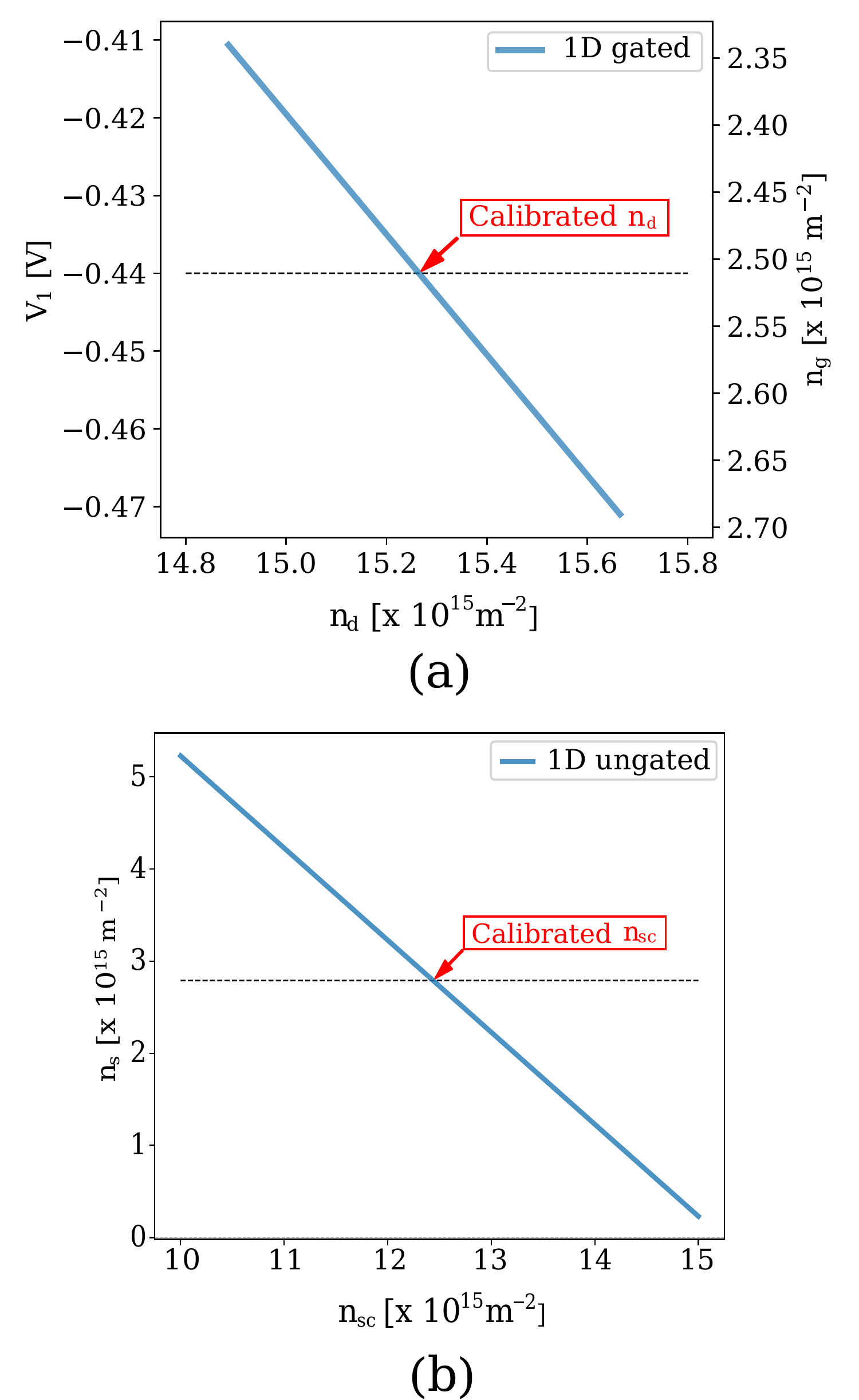}
    \caption{Illustration of the model calibration for sample A1a. Panel (a) shows the  pinch-off $V_\mathrm{1}$ (left axis) or equivalently $n_\mathrm{g}$ (right axis) calculated with the gated 1D model as a function of doping density $n_\mathrm{d}$. The horizontal dashed line shows a typical experimental value of $V_\mathrm{1}$. Panel (b) shows $n_\mathrm{s}$ as a function of the surface charge density $n_{\rm sc}$ using the 1D ungated model. The value of $n_\mathrm{d}$ is set to the intersection of the blue and dashed lines of panel (a). The horizontal dashed line shows the value of $n_{\rm bulk}$ and its intersection with the simulation result shows the calibrated value of $n_{\rm sc}$.}
    \label{fig:V1D_ndop_sc}
\end{figure}

Our model has two free parameters. One is the dopant density $n_\mathrm{d}$. It sets the 2DEG charge density underneath the ``gated" region at $V_\mathrm{g}$ = 0 equal to $n_\mathrm{g}$. The second is the surface charge density $n_{\rm sc}$. It sets the 2DEG density underneath the ``ungated" region, $n_\mathrm{s}$, for a given $\rm n_g$. Our model allows for a spatially varying density even in the absence of applied voltage, \textit{i.e.} $n_\mathrm{s}\ \ne\ n_\mathrm{g}$. As we shall see in section \ref{sec:discussion}, there are multiple experimental evidences that point towards the fact that these two densities are in fact equal ($n_\mathrm{s}\ =\ n_\mathrm{g}$) due to ``Fermi level pinning" and the model could be further simplified. Our calibration always leads to $n_\mathrm{s}\ \approx\ n_\mathrm{g}$ within 10\% which is consistent with Fermi level pinning. 

To calibrate our model, we use a two step process and two experimental values, $V_\mathrm{1}$ and $n_{\rm bulk}$. First, we vary $n_\mathrm{d}$ and calculate the  pinch-off voltage $V_\mathrm{1}$ in the ``gated" region. We set $n_\mathrm{d}$ so that the simulated $V_\mathrm{1}$ matches the experimental value. This sets $n_\mathrm{g}$. In the second step, we vary $n_{\rm sc}$ and calculate the density $n_\mathrm{s}$ in the ungated region. We set $n_{\rm sc}$ so that $n_\mathrm{s}$ matches the experimental 2DEG bulk charge density $n_{\rm bulk}\rm = 2.79 \times 10^{15} m^{-2}$. The calibration process is illustrated in Figure \ref{fig:V1D_ndop_sc}(a) (first step) and Fig. \ref{fig:V1D_ndop_sc}(b) (second step). It is repeated for each QPC.

\begin{figure*}
    \centering
    \includegraphics[scale=0.45]{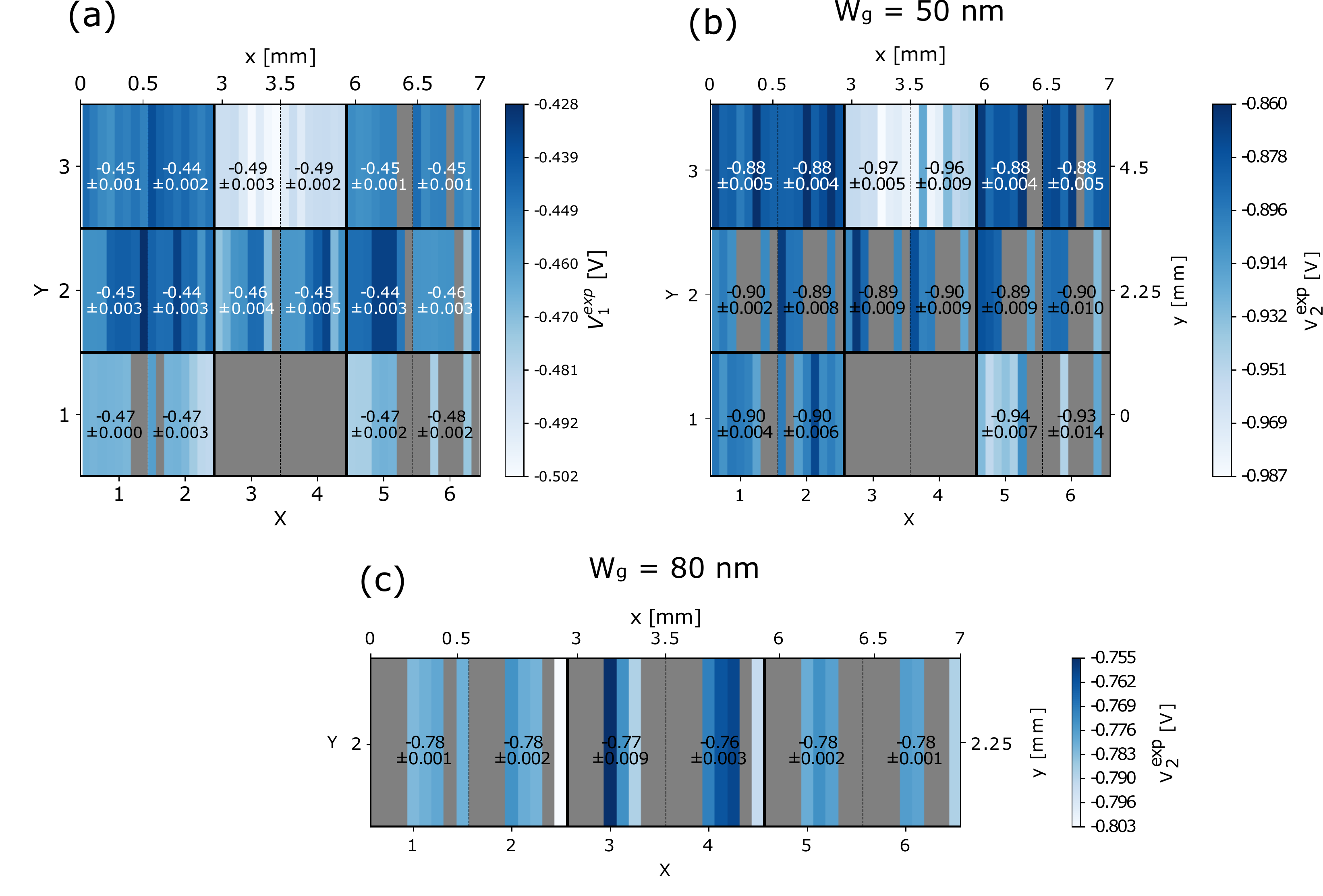}
    \caption{Colormaps of the variation of the experimental pinch-off values over the die. (a) ungated region values $V_{\rm 1}^{\rm exp}$; (b) and (c) narrow gate region values $V_{\rm 2}^{\rm exp}$ with $W$\textsubscript{g} = 50 nm and $W$\textsubscript{g} = 80 nm respectively. Each thin stripe corresponds to a different sample. For each set of QPCs, the average value and the standard deviation between the different samples of the set is also shown. The  top scale indicates the positions of the samples on the wafer in mm. The scale is not linear as consecutive pairs of saples are separated by 2 mm. No data is available in the gray regions (un-measured or distributed between (b) and (c).) }
    \label{fig:POexp}
\end{figure*}

Figure \ref{fig:POexp}(a) shows the variations of $V_\mathrm{1}$ for all the devices that have been measured. We find that the variations for QPCs within the same set are small, on the order of 0.5\% ($\pm 2$ mV). Therefore, using a unique average value of $n_\mathrm{d}$ to model a given set would give identical result with respect to the  QPC per QPC calibration. However, the variations of $V_\mathrm{1}$ for QPCs of different sets are larger - of the order of $\rm 10\%$ (40 mV). They are of the same order of magnitude as typical variations observed between different cooldowns. They imply the presence of significant variations over large distances of $\rm n_g$. We suspect that similar variations of $n_s$ are also present, see the discussion in section \ref{sec:discussion}. In the dices X = 3 and 4 with Y = 3, the calibration with $V_\mathrm{1}\approx -0.49$ V gives $n_\mathrm{g}\ =\ n_\mathrm{s}$ while in the other samples the two densities differ by less than 10\%.

\subsection{Simulations of the QPC regions pinch-off voltages V\textsubscript{3}}

After calibrating the model, we performed 3D simulations of the ``QPC" region to calculate the pinch-off voltage $V_\mathrm{3}$. Figure \ref{fig:A1B1_A1C1_sim_exp} shows the predicted (dashed lines) and measured (full lines) pinch-off voltages. We compare $V_\mathrm{3}$ as a function of $L$ (top panel, A samples), $R$ (middle panel, B samples) and $L$ (bottom panel, C samples).  Figure \ref{fig:A1B1_A1C1_sim_exp} highlights the main results of this article. It shows that the simulations correctly capture the pinch-off voltages. The main features of interest of Fig. \ref{fig:A1B1_A1C1_sim_exp} are:

(P1) Overall the simulations predict the pinch-off voltages quantitatively with a precision of the order of 10\%. 

(P2) There are significant experimental $V_\mathrm{3}$ variations in between QPCs with the same nominal characteristics. They are also of the order of 10\%. For instance the values of $V_\mathrm{3}$ observed for the four A2 samples (A2a, A2b, A2c and A2d) range from -2.2 V to -1.8 V, while the numerics predict a $V_\mathrm{3}$ close to -1.8 V. We also observed similar variations of the values of $V_\mathrm{3}$ (of the order of 0.1 V) on the same QPCs between different cooldown. Hence, the accuracy of the predictions is as good as the level of reproducibility of the experiments. Getting beyond this accuracy would involve a local in-situ calibration of the model so that any spatial variations of $n_\mathrm{s}$, $n_\mathrm{g}$ within the wafer would be accounted for. One could, for instance, include an additional QPC in the device, close to the active part of interest, and use the associated $V_\mathrm{3}$ value to calibrate the modeling with the actual local electronic density.

(P3) The $V_\mathrm{3}$ dependence on the QPC nominal characteristics $L$, $R$ and $W$ are correctly reproduced qualitatively. 

(P4) The predicted $V_\mathrm{3}$ is almost always smaller (in absolute value) than the experimental one by an offset of the order of 0.1--0.2 V. This indicates that our calibration slightly underestimates the value of the electronic density by 5--10\%. We attribute this fact to disorder as explained in section \ref{sec:disorder}.

\begin{figure}
    \centering
    \includegraphics[width=0.49\textwidth]{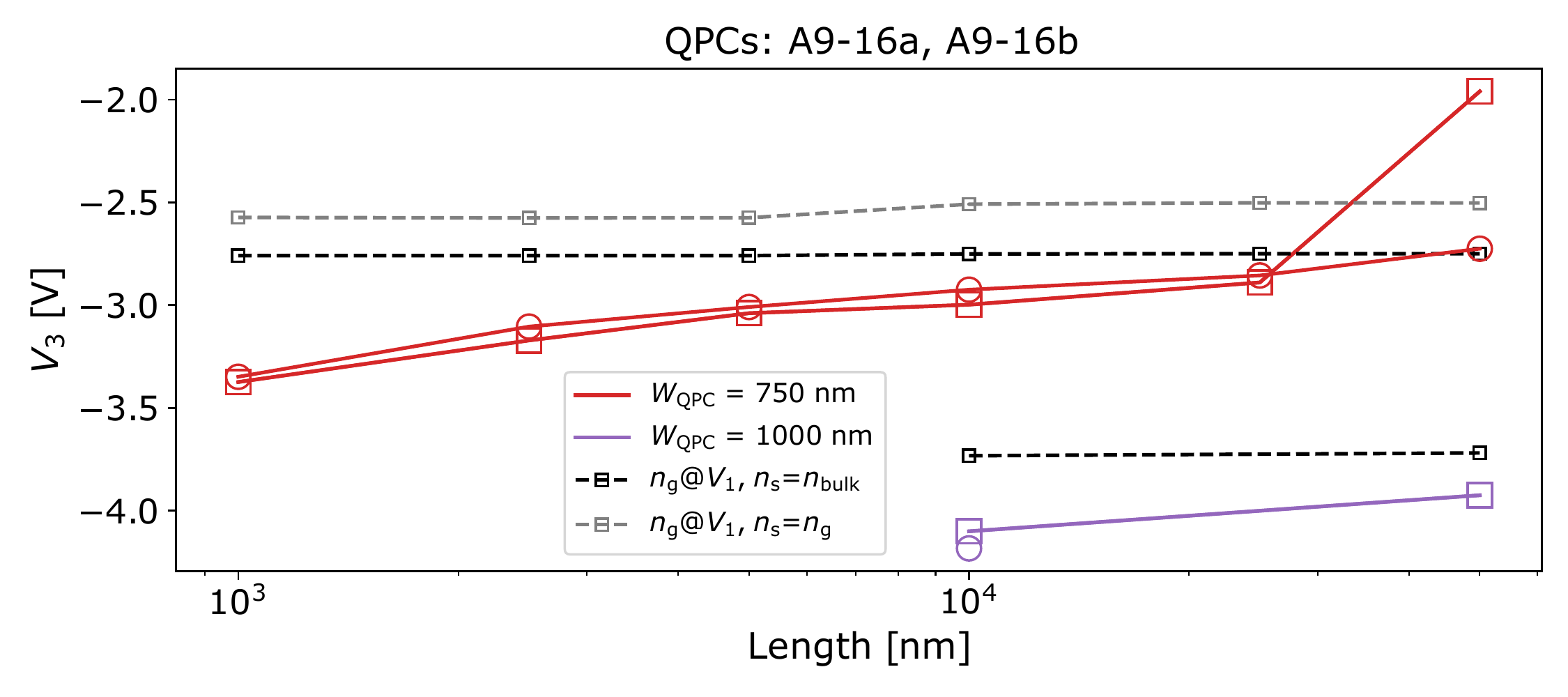}
    \caption{Experimental results of $V_\mathrm{3}$ for the large A designs. The dashed lines correspond to the simulation for an infinitely long sample. The different symbols correspond to sample a: ($\rm X_1Y_2$ , squares) and b ($\rm X_5Y_2$, circles). Two different calibrations have been used in the simulations: the one described in section \ref{sec:sim} (black dashed line) and a secondary calibration that enforces $n_\mathrm{s} = n_\mathrm{g}$ (gray dashed line).}
    \label{fig:A2_large_PO}
\end{figure}

Figure \ref{fig:A2_large_PO} shows the $V_\mathrm{3}$ data on samples with lengths 1 $\upmu \mathrm{m}\ \le L \le$ 50 $\upmu$m. For such long samples, the simulations predict that $V_\mathrm{3}$ should not depend on $L$. This trend is already observed in Fig. \ref{fig:A1B1_A1C1_sim_exp} for lengths exceeding 1 ${\upmu}$m. Indeed, the largest length scale in the problem is the distance between the 2DEG and the gate, \textit{i.e.} d $\approx$ 110 nm. When $L \gg$ d, $V_\mathrm{3}$ no longer depends on $L$. In practise, we have found that for $L \ge$ 5d, one has already reached the infinite $L$ limit in the simulations. Hence, the simulations for devices with $L \ge$ 5d are done by supposing $L=\infty$, \textit{i.e.} a system invariant by translation along the $y$ direction. We have used two different calibrations of the model: the same one as described in the preceding section (black) and a different one where we calibrate $n_\mathrm{g}$ with the experimental $V_\mathrm{1}$ and then set $n_\mathrm{s} = n_\mathrm{g}$. Both simulations give similar results and fail to capture the main experimental observation of Fig. \ref{fig:A2_large_PO} which is, 

(P5) $\mathrm{V_3}(L)$ has a large variation of $\approx$ +600 mV as the sample length goes from $L$ = 1 $\upmu$m to $L$ = 50 $\upmu$m (from -3.34 V at L =  1 $\upmu$m to -2.72 V at L =  50 $\upmu$m for $W_{\rm QPC}$ = 750 nm) .

Property (P5) cannot be explained by the model that we have used so far. In order to account for (P5), one must take into account the smooth density fluctuations that take place on long scales. Indeed, in the presence of spatial variations of the density along the $x$ direction, the pinch-off $V_\mathrm{3}$ is determined by the position in $x$ where the density is smallest. A model analyzing semi-quantitatively the role of the disorder will be presented in section \ref{sec:disorder}.

\subsection{Simulations of the narrow gate region pinch-off voltages $V_\mathrm{2}$}

\begin{figure}
     \centering
          \begin{subfigure}{0.45\textwidth}
         \centering
         \includegraphics[width=\textwidth]{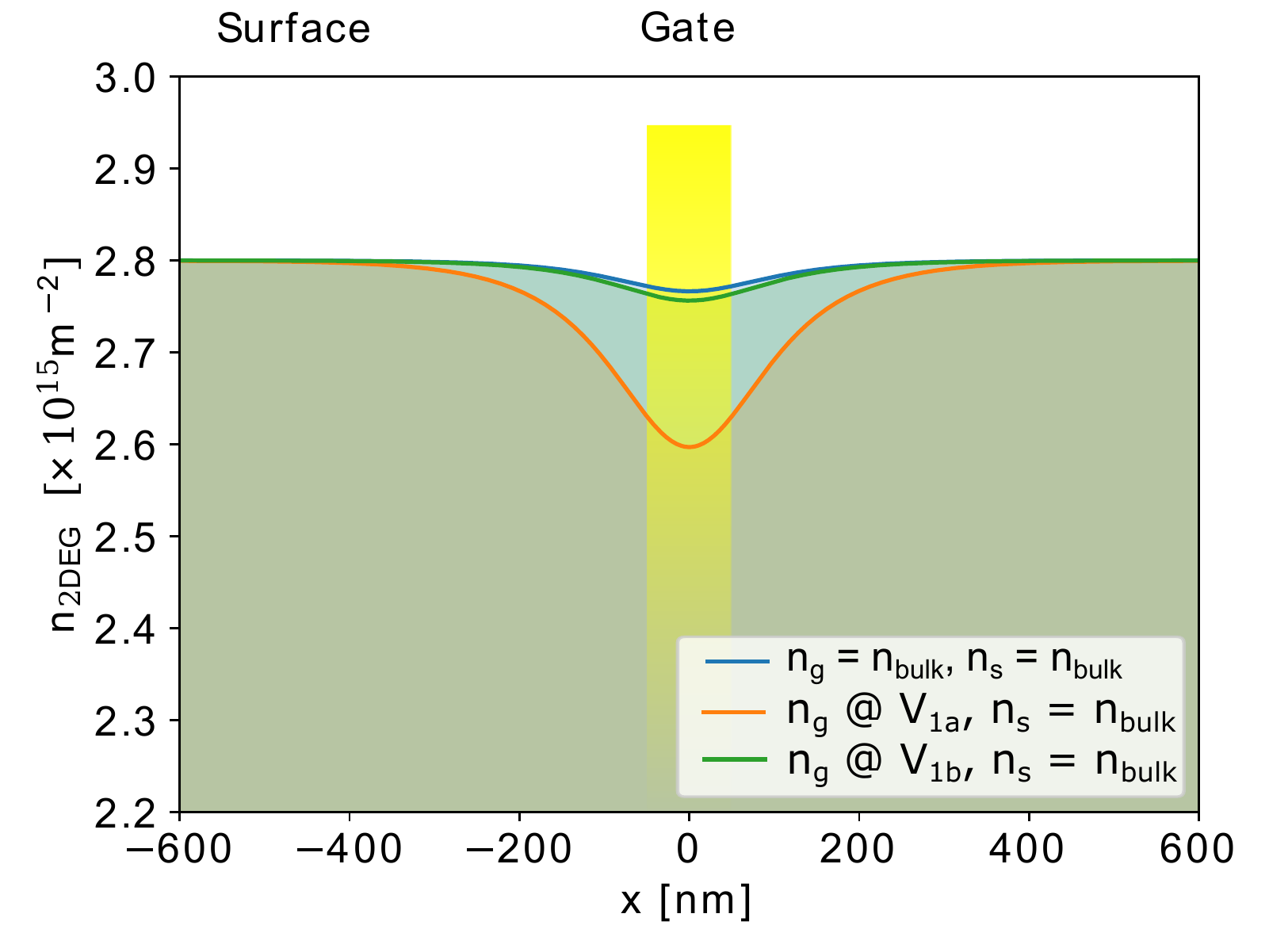}
         \caption{}
         \label{fig:50nm_n2DEG}
     \end{subfigure}
     \hfill
     \begin{subfigure}{0.45\textwidth}
         \centering
         \includegraphics[width=\textwidth]{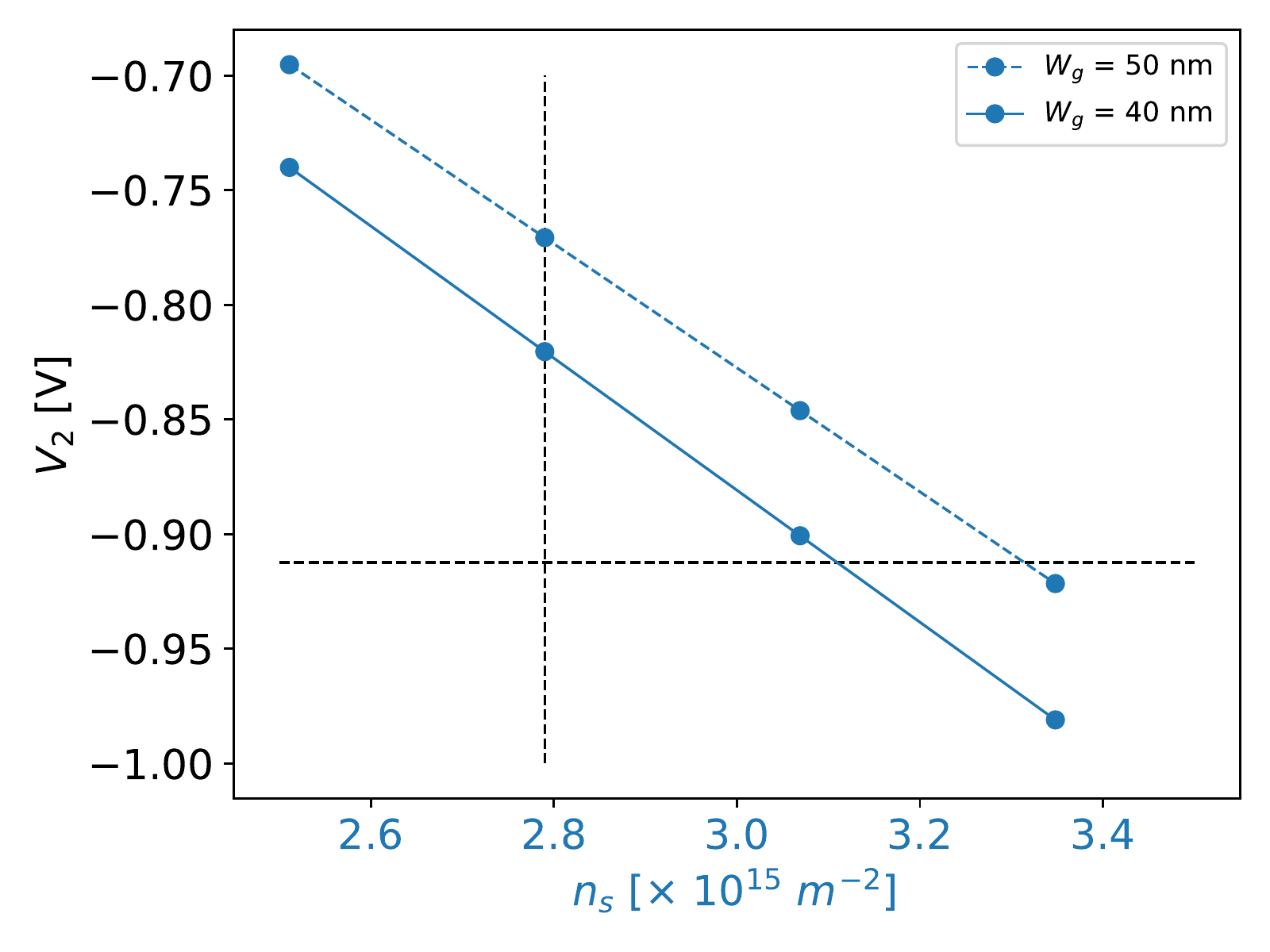}
         \caption{}
         \label{fig:Gatewidth_Vp_2D}
     \end{subfigure}
     \caption{(a) Simulated electronic density profile as a function of $x$ at $\rm V_g$ = 0, in the narrow gate region with $W_\mathrm{g}$ = 50 nm. The orange and green 2DEG profiles are the simulation results obtained from the standard calibration for QPC B2a and B2b. The blue line uses a calibration where $n_\mathrm{s}\ =\ n_\mathrm{g}\ =\ n_{\rm bulk}$. (b) $\rm V_2$ as a function of $n_\mathrm{s}\ =\ n_\mathrm{g}$ for two different gate widths W = 50 nm (dashed blue line) and $W$ = 40 nm (full blue lines). The vertical line shows $n_\mathrm{s}\ =\ n_{\rm bulk}$ and horizontal line shows the experimental value of $V_\mathrm{2}$.}
     \label{fig:2D_res}
\end{figure}

We now turn to the simulations of the ``narrow gate" region. They correspond to the very long ($>$ 20 $\upmu$m) but thin (50 nm wide) gate. Figure \ref{fig:50nm_n2DEG} shows a typical simulation of the electronic density versus $x$ at zero applied gate voltage. The different curves correspond to different densities of surface charge and dopants. Specifically, we have used different calibrations for the density $n_\mathrm{s}$ far away from the gate and the density $n_\mathrm{g}$ under a  (wide) gate. While the simulated 2DEG density varies below the gate, this variation is smaller than 5\% which corresponds to the variation of $V_\mathrm{1}$ that we have observed on different samples. It follows that our findings are fully compatible with a uniform density at zero voltage. 

The main observation we make for the voltage $V_\mathrm{2}$ is that our predictions are significantly lower than the experimental data for both values of width $W_\mathrm{g}$, see Table \ref{tab:V2_std}. More precisely:

(P6) The simulations systematically underestimate the magnitude of $V_\mathrm{2}$ by $\approx 0.15$ V (20\%).

\begin{table}
\begin{center}
\begin{tabular}{| c | c | c | c | }
\hline
\rowcolor{lightgray}  & $V_2$ exp. & $\sigma_{V_2}$ exp. &  $V_{2}$ sim. mean \\
50 nm & -0.90 & 0.032 & -0.73 \\ 
80 nm & -0.77 & 0.010 & -0.63 \\
\hline
\end{tabular}
\caption{\label{tab:V2_std} Comparison between experimental (exp) and simulated (sim) values of $V_\mathrm{2}$ for the two different widths $W$\textsubscript{g}. $\sigma_{V_\mathrm{2}}$ is the standard deviation of $V_\mathrm{2}$ between different QPC. We observe a systematic deviation of $\approx\rm 0.15 V$ (20\%) between the simulation and the experiments.}
\end{center}
\end{table}

The error in (P6) is the largest discrepancy we have observed between the simulations and the experiments. We identify four possible origins for this discrepancy. (1) The bulk value $n_\mathrm{s}$ is higher than the one we used. (2) The width $W_\mathrm{g}$ is narrower than what is drawn in the design. Gate fabrication uses standard lithographic technique with e-beam insolation of a resist, chemical lift-off of the resist followed by metal deposition and chemical lidt-off of the residual resist. This process should have an accuracy better than $10$ nm in the width of the gates. (3) The width $W_\mathrm{g}$ fluctuates along the gate due to lithography accidents. (4) There are density fluctuations of the 2DEG due to disorder. 

Figure \ref{fig:Gatewidth_Vp_2D} shows the predicted value of $V$\textsubscript{2} as a function of the 2DEG density $n_\mathrm{s}$ assuming a uniform 2DEG density at $V$\textsubscript{g} = 0  ($n_\mathrm{s}\ =\ n_\mathrm{g}$). The vertical line corresponds to the nominal value $n_\mathrm{s}\rm = 2.8 \cdot 10^{15} m^{-2}$. The horizontal line is the measured value of $V_\mathrm{2}$. We see that to obtain the experimental value of $V_\mathrm{2}$ for $W_\mathrm{g}\ =\ 50$ nm, one needs $n_\mathrm{s}\rm =\ 3.4\cdot 10^{15}\ m^{-2}$ which is unreasonably high (This 21.5\% higher than the nominal value while typical density variations inside a wafer are in the 5--10\% range). Hence, we can rule out (1) as the origin of (P6). The straight blue line in Figure \ref{fig:Gatewidth_Vp_2D} shows the value of $V_\mathrm{2}$ obtained when one reduces the width of the gate by 20\%, \textit{i.e.} $W_\mathrm{g}$ = 40 nm. We see that this is not sufficient to reproduce the experimental data and larger variations of $\rm W_g$ would be visible on the SEM images. In contrast the SEM images indicate a width that is  slightly larger than 50 nm. Hence, we rule out (2). Finally, we do not observe sample to sample variations of $V_\mathrm{2}$ and the SEM pictures do not show fluctuations of the width $W_\mathrm{g}$ along the gate. Hence we rule out (3). 

The last scenario (4) corresponds to smooth spatial fluctuations of the density inside the sample. This could be due to \textit{e.g.} doping density or background doping fluctuations \cite{zhou2015,qian2017,chung2019}. Indeed, if the electronic density varies underneath the 20 $\upmu$m long gate, the corresponding $V_\mathrm{2}$ pinch-off will be given by the region of {\it largest} density. This interpretation is fully consistent with the observation of the significantly large sample to sample fluctuations of $V_\mathrm{3}$. In section \ref{sec:disorder} we perform a systematic analysis of the effect of long range disorder on $V_\mathrm{1}$, $V_\mathrm{2}$ and $V_\mathrm{3}$. We find that a 5--10\% density fluctuation consistently explain (P2), (P5) and (P6).

\section{Critical discussion of the modeling}
\label{sec:discussion}

In this section we discuss various aspects of the modeling in more detail. We emphasize again that we refrained from the particularly difficult goal of trying to predict the bulk density of the device. That is, to develop a model capturing the microscopic details along the 1D $z$-direction of our samples.  While the corresponding physics is well understood and has been studied rather extensively, the resulting electronic density depends on many parameters which are often poorly known. These microscopic parameters include the density of dopants, the fraction of dopants that are ionized (or equivalently the precise dopant ionization energies - including the so called DX centers), the residual doping in the bulk of the wafer, the density of surface charges (or equivalently the precise value of the binding energy of the surface states), the workfunction of the metals used in the electrostatic gate with respect to GaAs, the values of the band offsets, the effective masses, the relative dielectric constants of the different materials \cite{Buks1994, Chung2017, Davies1997, Weisbuch1991} etc. Making quantitative predictive simulations with so many unknown parameters that depend on the growth condition of the wafer is very challenging. It also serves a very different purpose, more related to wafer characterization than to the understanding of the devices made out of it.

Our goal instead is to be able to predict the spatial variations along the 2D $x$- and $y$-directions. We  use experimental measurements to tabulate the result of the interplay between all the above mentioned parameters. Indeed, and this is a very important point, while this interplay is quite subtle at room temperature, at sub-Kelvin temperatures on the other hand all the possible source of charges (surface, dopants) are essentially frozen. Hence, while they do contribute to the electronic density, their effect boils down to a contribution to the 2DEG electronic density that can be measured independently through \textit{e.g.} Hall measurements. The fact that the charge sources are frozen, a well established experimental fact, coupled to the linearity of the Poisson equation, means that to predict the effect of the gate voltages on the 2DEG density one only needs: (i) the distance of the 2DEG with respect to the gates and (ii) the low temperature 2DEG density profile in the $xy$ plane at zero applied gate voltage. This is precisely what we are trying to capture in our simulations. Below we discuss how the choices of (i) and (ii) made in our modeling affect the results.

\subsection{Role of quantum capacitance and quantum fluctuations on the electronic density}
\label{sec:min}

Let us discuss a 1D minimum model to discuss the electronic density in the ``ungated" (bulk) or ``gated" regions. These regions are sufficiently large so that the 2DEG can be considered far from the gate boundaries. The spatial variation of the 2DEG density in the $xy$ plane can thus be ignored, assuming no disorder. We are left with a, possibly complex, 1D problem along the $z$-direction. We describe the 2DEG by its 2D density of states $\rho$. We suppose the vertical distance to the gate (at voltage $V_\mathrm{g}$) to be d. Lastly, we assume there is an arbitrary distribution of doping charges $n_\mathrm{d}(z)$. It includes the ionized dopants, the surface charge and any other frozen charge that might be present in the system. The 2DEG density $n_\mathrm{s}$ is given by $n_\mathrm{s}=\rho \mu$ where $\mu$ is the chemical potential of the 2DEG. Assuming without loss of generality, that the 2DEG is grounded, the electrochemical potential vanishes so that $\mu -eU(z=0)=0$. The model reduces to solving the Poisson equation,

\begin{equation}
\frac{\partial^2}{\partial z^2} U(z) = \frac{e}{\epsilon} n_{\rm d}(z)
\end{equation}

with the boundary conditions $\rm U(d) = V_\mathrm{g}$ and $\rm \partial_\mathrm{z} U(0) = (e^2\rho/\epsilon) U(0)$. This equation being linear, its solution can be written as a linear combination of two terms $\rm U(z) = U(z,V_\mathrm{g}=0) + V_\mathrm{g} U(z,V_\mathrm{g}=1)$. We thus arrive at,

\begin{equation}
\left(\frac{1}{e^2\rho} + \frac{d}{\epsilon}\right)e n_\mathrm{g} = V_{\rm 1} - V_\mathrm{g}
\label{eq:Surface1D}
\end{equation}

where the parameter $V_\mathrm{1}$ is the pinch-off voltage. It corresponds to the contribution of $n_\mathrm{d}(z)$ to the electronic density. In such a simple model, $V_\mathrm{1}$ could be expressed explicitly in terms of $n_\mathrm{d}(z)$. However, we will refrain from doing so and take it as an experimentally measurable parameter. Equation (\ref{eq:Surface1D}) provides a direct conversion relation between 2DEG density to voltages. For a distance of d = 100 nm and using $\epsilon=12\epsilon_\mathrm{0}$, the density is $n_\mathrm{g}\rm = 6.6\cdot 10^{15} m^{-2} V_{\rm 1}$. For our stack, \textit{i.e.} a bulk density of $n_\mathrm{s} = 2.8 \cdot 10^{15}$ m$^{-2}$ and d = 110 nm, we calculate $V_\mathrm{1}$ = (2.8/6.6)*(110/100) = 0.46 V. The latter is the predicted pinch-off voltage in the ``gated" region. This simple calculation actually matches the measured value of $V_\mathrm{1}$.

In the calculation above we have neglected the contribution from the density of states. Indeed, the contribution of the 1/($e^2\rho$) term, \textit{i.e.} the inverse of the quantum capacitance, is for most devices negligible compared to the inverse of the geometrical capacitance $d$/$\epsilon$. For the QPCs studied in this paper, the quantum capacitance term adds a correction of 2\% to the voltage pinch off. The latter is estimated using the effective mass approximation and assuming only the first sub-band as occupied when calculating $\rho$. That is $\rho = \frac{m^*}{\hbar^2\pi}$ with $m^*\approx 0.067 m_\mathrm{0}$ for GaAs. A 2\% correction is smaller than our experimental resolution. We conclude that the various pinch-off voltages are almost entirely controlled by the electrostatics of the problem, \textit{i.e.} the geometrical capacitance. They are enough to characterize the distribution of charges in the 2DEG. 

 The value d = 110 nm is the physical distance between the electrostatic gate and the GaAs/AlGaAs interface. In principle one should take into account the finite width of the 2DEG which is of the order of $10$ nm. This effect is partially taken into account in the simulations at the Thomas-Fermi level, but would be more pronounced if the quantum fluctuations along the $z$-direction were included.
 We have performed various full self-consistent 1D Schrodinger-quantum simulations (not shown). They show a
 small correction of the final width of the 2DEG of less than 1\%. 

\subsection{Fermi level pinning of the dopants at room temperature} 
\label{sec:fermi}

In our model we have assumed that, at $V_{\rm g} = 0$ the density underneath a gate could be different from the density away from a gate. This is reflected in the presence of the two parameters $n_{\rm s}$ (density in the ungated region) and $n_{\rm g}$ (density in the gated region) that can {\it a priori} take distinct values $n_{\rm s} \ne n_{\rm g}$. We have found {\it a posteriori} that the experiments are best fitted by $n_{\rm s} \approx n_{\rm g}$ indicating  that the $V_{\rm g} = 0$ density has small spatial variations inside a given sample. Here, we discuss the phenomena of ``Fermi level pinning" in the dopant region at room temperature, i.e. the possibility that --at room temperature-- the dopant layer behaves essentially as a (metallic like) equipotential. The existence of Fermi level pinning in an actual stack requires a sufficently high concentration of dopant and sufficiently low disorder in the dopant region for the dopants to remain on the matallic side of the metal-insulator transition at room temperature (however so slightly). The presence of Fermi level pinning would imply $n_{\rm s} = n_{\rm g}$. We argue that there are strong experimental evidences for Fermi level pinning in our samples. This fact can be used to reduce the model to a single fitting parameter. 

A homogeneous dopant distribution leads naturally to $n_\mathrm{s}\ \ne\ n_\mathrm{g}$ unless the surface charge density accidentally matches the contribution coming from the workfunction at the gate-GaAs interface. An opposite hypothesis --- Fermi level pinning --- is that, at room temperature, the {\it electric potential} (not the ionized dopant density) is constant inside the dopant layer. Fermi level pinning happens when the charges in the dopant layer are sufficiently mobile to form a metallic-like equipotential. The associated charge distribution (now spatially dependent) gets frozen upon cooling the sample to low temperature. By construction, Fermi level pinning implies $n_{\rm s} = n_{\rm g}$ since the sources of spatial inhomogeneities (the gates) are situated above the dopant region, hence screened. Below, we list experimental evidences for the presence of Fermi level pinning in GaAs/AlGaAs heterostructures. These evidences are very strong for {\it some} heterostructures (in particular in the case of ``delta doping" where the dopant concentration is very high hence more likely to form a band) but it is not certain that Fermi level pinning is present in {\it all} of them.

\begin{itemize}
\item In \cite{Buks1994,Buks1994b}, the field effect of HEMTs made in these heterostructures is shown to disappear at high temperature, indicating that the dopant layer screens the effect of the gate. 
\item Fermi level pinning is also the natural explanation for the hysteretic effect known as ``bias cooling" \cite{Buks1994b}: when one applies a voltage on an electrostatic gate during the cooling of the sample, one observes that the low temperature current versus gate voltage characteristics gets shifted horizontally by the same amount. For instance if a sample normally pinches at $-1.5$ V for a regular cooling, a $+1$ V bias cooling will make it pinch at $-0.5$ V. This is a strong indication that the voltage applied at room temperature did not affect the electronic density. Upon cooling, the dopants get frozen and must be considered as a fixed charge density. Hence, to deplete the gas, what matters is the {\it variation} of the voltage with respect to the value used during the cooling, not the absolute value of the voltage. Bias cooling is often used by experimentalists to reduce the pinch-off voltage and avoid leakages. It has been observed repeatedly including in wafers nominally identical to the one used in the
experiments presented in this article. 
\item There are multiple experimental evidences showing that using different metals for the electrostatic gate, say gold and aluminium, give devices with very similar properties in terms of pinch-off voltages  \cite{FredPierre}. This is an additional experimental evidence for Fermi level pinning. Indeed, different gate materials, such as gold and aluminium, have very different work functions of the order of 0.8 V. In the absence of Fermi level pinning, one would get very different pinch-off values as well as signature of a strongly varying spatial distribution of the 2DEG density (visible in \textit{e.g.} quantum Hall effect experiments). None of these effects are observed experimentally.
\item We have found that the best fit to our model implies $n_{\rm s} \approx n_{\rm g}$ within 5\% which is unlikely
to happen accidentally.
\end{itemize}

We conclude that Fermi level pinning of the dopants is very likely present in our samples. This could be used to further simplify our model to a single parameter $n_{\rm s} = n_{\rm g}$ that can be calibrated in situ using pinch-off voltage.

We note that the phenomena of Fermi-level pinning could also be discussed with respect to the {\it surface states}. Our calculations show that such an effect, if present, could not account for phenomena such as the hysteresis observed in bias cooling but at most to half of the observed effect. The presence of Fermi-level pinning of the surface states would, however, further contribute to enforce $n_{\rm s} = n_{\rm g}$.

\subsection{Long range disorder and density fluctuations.}
\label{sec:disorder}

\begin{figure}
\centering
\includegraphics[width=0.47\textwidth]{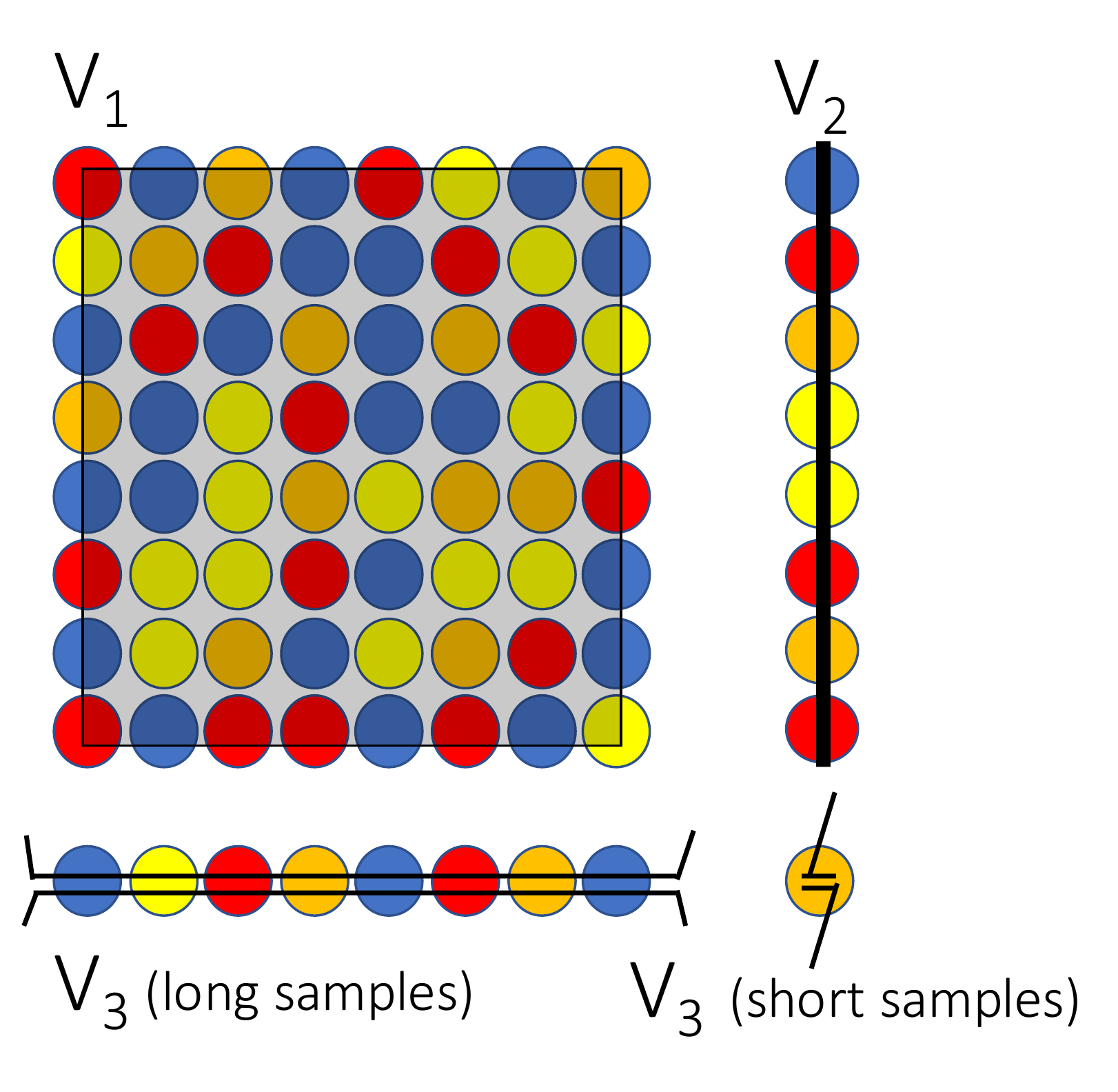}
\caption{Schematic of the percolation model used to explain the effect of the smooth long range disorder. Each circle corresponds to a region of size $\xi$ with a random electronic density (symbolized by different colors). As one increases the gate voltage towards negative values, the density decreases everywhere. The pinch-off is obtained when there is no path left with finite density to go from left to right. The fluctuations of density manifest themselves differently for the different pinch-off: 2D percolation problem for $V_{\rm 1}$ (upper left), many regions in parallel for $V_{\rm 2}$ (upper right) or in series for $V_{\rm 3}$ of long samples (lower left) while the fluctuations of density induce fluctuations of $ V_{\rm 3}$ for short samples (lower right). }
\label{fig:percolation}
\end{figure}

We end this article with a discussion of the role of disorder in the system. So far, all the analysis has been done assuming a perfect 2DEG whose spatial density profile is only affected by the electrostatic gates. Despite the very high mobility of the 2DEG, there remains some disorder in the system. There are several types of disorder that can be present in the system\cite{Buks1994,Buks1994b} including inhomogeneities in the dopant density, interface roughness or background impurities. Note that some types of disorder such as interface roughness may very well affect the conductance. However as interface roughness varies mostly on short (atomic) scale it is unlikely to significantly affect the electronic density unless the disorder is very strong. This type of disorder can be ignored for the purpose of this discussion. Indeed, the goal of this section is to understand the effect of disorder on the pinch-off properties of the device. Hence we focus on the slowly varying part of the disorder on scale larger than the Fermi wave length. We attribute this disorder mostly to variations of dopant density, but other sources could be present as well without affecting the discussion that follows. Recent experimental works that explicitly study the spatial variation of the electronic density include \cite{zhou2015,qian2017,chung2019}
 
We construct a simple percolation model to discuss the effect of long range disorder on the three thresholds $V_\mathrm{1}$, $V_\mathrm{2}$ and $V_\mathrm{3}$. A schematic of the model is shown in Fig.\ref{fig:percolation}. We will see that this simple model can account for all the systematic discrepancies observed between the simulations and the experiments, at least qualitatively. Let's consider a $L_x \times L_y$ 2DEG sample. We suppose that the density is slowly varying on a typical length scale $\xi$, the disorder correlation length. The 2DEG can thus be considered as made of $L_x/\xi \times L_y/\xi$ small samples (the circles in Fig.\ref{fig:percolation}), hereafter referred to as ``cells". Typically, we expect $\xi$ to be of the order of a few hundred nanometers for a disorder due to dopant density fluctuations. Each cell has a constant density $\rm n_{ij}$ with i $\in\{1....L_x/\xi\}$ and j $\in\{1....L_y/\xi\}$. The value of the density $\rm n_{ij}$ in cell $\rm (i,j)$ is a random variable of mean $n_\mathrm{g}$ and variance $\rm\sigma^2_g$, independent from the density in other cells. For definiteness, we suppose that the associated probability density is flat,

\begin{eqnarray}\label{eq:distrib}
P(n<n_{\rm ij}<n+dn) =  \nonumber\\
\frac{1}{2\sqrt{3}\sigma_\mathrm{g}}\theta(n-n_\mathrm{g}-\sqrt{3}\sigma_\mathrm{g})\theta(n_\mathrm{g}+\sqrt{3}\sigma_\mathrm{g} - n) dn
\end{eqnarray}   

where $\theta(x)$ is the Heaviside function. Last, we suppose that the pinch-off voltage on each cell $\rm (i,j)$ is simply proportional to $\rm n_{ij}$ as found in the minimal model of section \ref{sec:min}. 

Let us first examine the implication of this model for the threshold $V_\mathrm{1}$. In the absence of density fluctuations, the conductance through the gated region vanishes when the gate voltage depletes the 2DEG entirely. In presence of fluctuations, however, depleting only a fraction of the 2DEG cells suffice. That is, if the fraction p of remaining cells with non zero density is bellow the percolation threshold $p_c \approx 0.6$ of the 2D square lattice of cells, then the conductance vanishes. We introduce the probability $P\rm(n_{ij}\ge n)$ for a cell to have a density larger than $n = n_\mathrm{g} + \delta n$,

\begin{equation}
P(n_{ij} \ge n_\mathrm{g} + \delta n) = \frac{1}{2} - \frac{\delta n}{2\sqrt{3}\sigma_\mathrm{g}}
\end{equation}

for $|\delta n| \le \sqrt{3} \sigma_\mathrm{g}$. The percolation threshold corresponds to $P(n_{\rm ij}\ge n_\mathrm{g} +\delta n) = p_\mathrm{c}$. From the latter we obtain the variation $\delta V_\mathrm{1}$ induced by the density fluctuations,

\begin{equation}
\label{eq:disorderV1}
\frac{\delta V_\mathrm{1}}{|V_\mathrm{1}(\sigma_\mathrm{g}=0)|} = (2 p_\mathrm{c} - 1)\sqrt{3}\frac{\sigma_\mathrm{g}}{n_\mathrm{g}} \approx 0.34 
\frac{\sigma_\mathrm{g}}{n_\mathrm{g}}
\end{equation}

Equation (\ref{eq:disorderV1}) leads to a positive variation of $V_\mathrm{1}$. Since $V_\mathrm{1}<0$, it leads to a decrease of $V_\mathrm{1}$ in absolute value. Conversely, not taking the density fluctuations into account when estimating $V_\mathrm{1}$ leads us to underestimate the density $n_\mathrm{g}$.

Next, we examine the implications of the disorder model for the threshold $V_\mathrm{2}$. In the narrow gate region, $L_x$ is smaller than the correlation length $\xi$. Therefore, the current travels through $L_y/\xi$ cells in parallel. The pinch-off is reached when the voltage is sufficiently negative to deplete {\it all} the cells. Hence $V_\mathrm{2}$ corresponds to the voltage needed to deplete the cell with largest density. The probability for the cell with highest density to have a density smaller than $n_\mathrm{g} + \delta n$ is given by $[\mathrm{1/2} + \delta n/(2\sqrt{3}\sigma_\mathrm{g})]^{L_y/\xi}$. It corresponds to the probability that all cells have a density smaller than $n_\mathrm{g} + \delta n$. For $L_y \approx 50\upmu$m $\gg \xi$ this probability is strongly peaked around $\delta n = \sqrt{3} \sigma_\mathrm{g}$, from which we obtain the variation $\delta V_\mathrm{2}$ induced by the density fluctuations,

\begin{equation}
\label{eq:disorderV2}
\frac{\delta V_2}{|V_2(\sigma_\mathrm{g}=0)|} = - \sqrt{3}\frac{\sigma_\mathrm{g}}{n_\mathrm{g}} \approx -1.7 \frac{\sigma_\mathrm{g}}{n_\mathrm{g}}
\end{equation}

The effect of disorder on $V_\mathrm{2}$ is around 5 times larger than on $V_\mathrm{1}$. It is also of opposite sign.
We can calculate the standard deviation $\sigma_{V_\mathrm{2}}$ of $V_\mathrm{2}$ due to sample to sample variations. We find,

\begin{equation}
\label{eq:disorderV2b}
\frac{\sigma_{V_\mathrm{2}}}{|V_\mathrm{2}(\sigma_\mathrm{g}=0)|} = 2 \sqrt{3}\frac{\xi}{L_y}\frac{\sigma_\mathrm{g}}{n_\mathrm{g}}
\end{equation} 

Last, we look at the influence of disorder on $V_\mathrm{3}$ in two different limits. For the small samples $L\le 2\ \upmu$m, the QPC region corresponds essentially to a single cell. In that limit, the fluctuations $\sigma_{V_\mathrm{3}}$ of the threshold $V_\mathrm{3}$ are simply given by the density fluctuations of a single cell and,

\begin{equation}
\label{eq:disorderV3}
\frac{\sigma_{V_3}}{|V_3(\sigma_g=0)|} = \frac{\sigma_\mathrm{g}}{n_\mathrm{g}}.
\end{equation} 

A second interesting limit corresponds to the very long samples 10 $\upmu$m $\le$ L $\le$ 50 $\upmu$m.  These samples correspond to the dual situation to $V_\mathrm{2}$: the different $L_x/\xi$ cells are in parallel instead of being in series. Therefore the pinch-off is limited by the cell that has the smallest density. The probability for the smallest density to be larger than  $n_\mathrm{g} + \delta n$ is given by $[1/2 - \delta \mathrm{n}/(2\sqrt{3}\sigma_\mathrm{g})]^{L_y/\xi}$. For $L_x \gg \xi$, we get,

\begin{equation}
\label{eq:disorderV3b}
\frac{\delta V_\mathrm{3}}{|V_\mathrm{3}(\sigma_\mathrm{g}=0)|} = \sqrt{3}\frac{\sigma_\mathrm{g}}{n_\mathrm{g}} \approx 1.7 \frac{\sigma_\mathrm{g}}{n_\mathrm{g}}
\end{equation}
\textit{i.e.} the fluctuations make it easier to pinch-off a long wire. This ends our analysis. Note that the precise value of the prefactors in Eqs.(\ref{eq:disorderV1}),
(\ref{eq:disorderV2}),(\ref{eq:disorderV2b}) and (\ref{eq:disorderV3}) depend on the choice of distribution Eq.(\ref{eq:distrib}) so that the percolation model should be used for trends, not
precised comparisons. 

\subsection{Comparison between the experiments and the percolation model}
\label{sec:accuracy}

Let's now go back to the experimental data and show that the above percolation model accounts for all the imperfections of the no-disorder model at a semi-quantitative level. The largest discrepancy between our predictions and the experiments is the one of $V_\mathrm{2}$ (property P6). Indeed, the simulations for perfect systems systematically show values of $V_\mathrm{2}$ that are around 20\% smaller (in absolute value) than what is observed experimentally. To account for this $\delta V_\mathrm{2}/|V_\mathrm{2}|\approx 0.2$, equation (\ref{eq:disorderV2}) implies that density fluctuations with $\sigma_\mathrm{g}/n_\mathrm{g} \approx 0.12$ occur in the system. Density fluctuations of 12\% is compatible with what is commonly believed by the community for this system if somewhat large \cite{zhou2015,qian2017,chung2019}. It is also compatible with what we have observed on larger scales on the fluctuations of $\rm V_1$ (see Fig. \ref{fig:POexp}) . Equation (\ref{eq:disorderV3}) then implies that the sample to sample variations of $V_\mathrm{3}$ are also of the order of 12\%. While we do not have enough statistics to properly estimate the variance of $V_\mathrm{3}$, a rough estimate from our data is of the order of 6\% (property P2, see Fig.\ref{fig:A1B1_A1C1_sim_exp}). 

\begin{figure}
    \centering
    \includegraphics[width=0.49\textwidth]{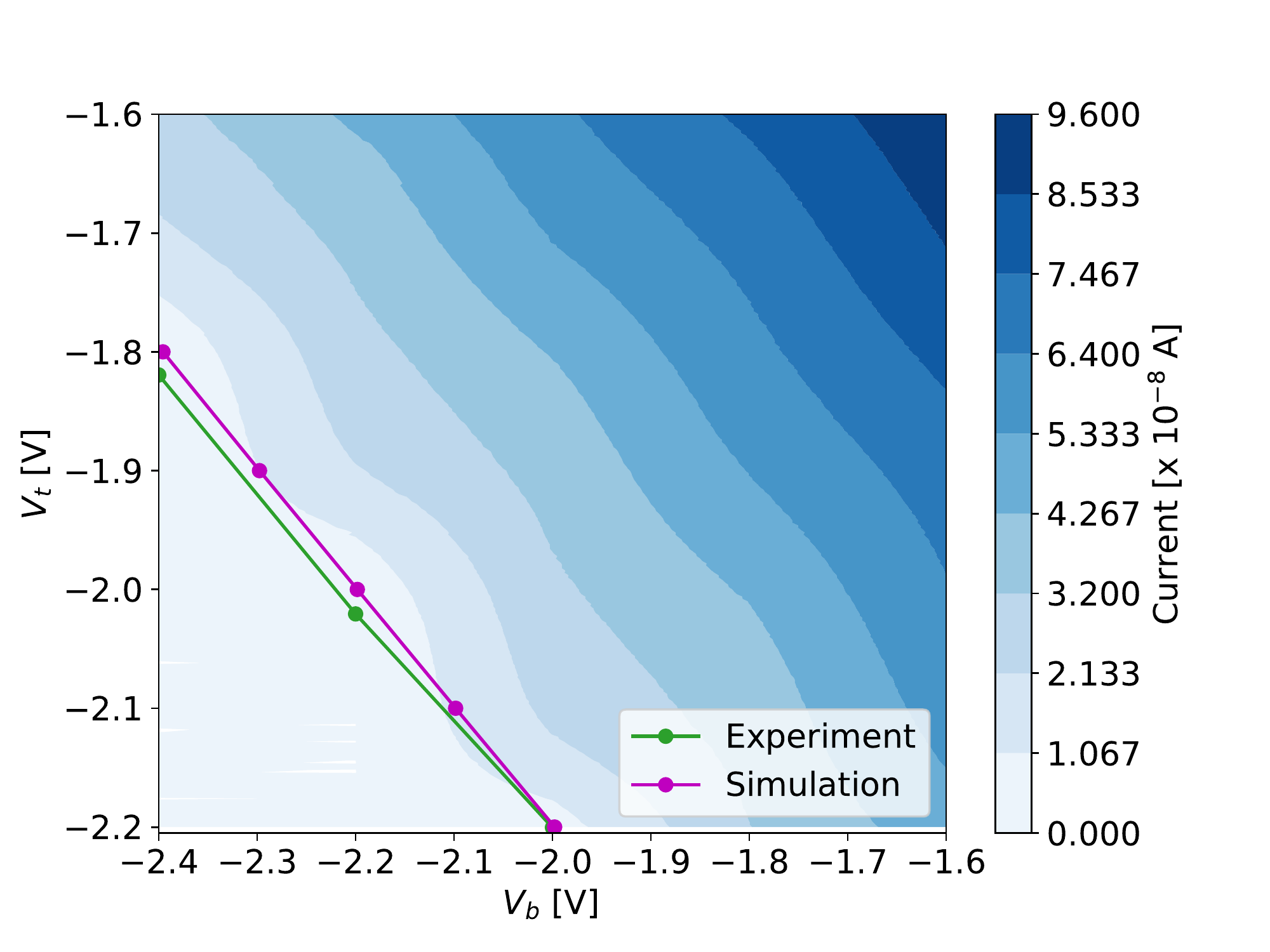}
    \caption{Colormap of the current versus the top ($V_\mathrm{t}$) and bottom ($V_\mathrm{b}$) gates that were biased separately for QPC A1a. Green (blue) lines show the experimental (simulated) pinch-off voltage. $V_\mathrm{3}\rm \approx -2.1$ V corresponds to $V_\mathrm{t}=V_\mathrm{b}$ and is used to calibrate the simulations in a single parameter model
$\rm n_g=n_s \approx n_{bulk} +5\%$}
    \label{fig:map_2D}
\end{figure}

Equation (\ref{eq:disorderV1}) then predicts a correction to $V_\mathrm{1}$ of 4\%. Taking that correction into account in our calibration would bring all our predictions in Fig.\ref{fig:A1B1_A1C1_sim_exp} down by 4\% (80 mV). This would significantly improve the match between experiments and simulations, see the discussion of property (P4). Another possible source of error in $V_\mathrm{1}$ stems from an imprecision when extracting the experimental value from the conductance curve. Near $V_\mathrm{1}$ the ``gated" region contribution to the overall conductance is much smaller than that of the ``narrow gate" region. The latter contribution thus obscures the conductance due to the ``gated" region. This adds an error to the extracted value of $V_\mathrm{1}$ that is not accounted by our theoretical model. 

Using Equation (\ref{eq:disorderV2b}), the small observed fluctuations $\sigma_{V_\mathrm{2}} \approx 5$ mV imply a correlation length $\xi \approx$ 1-2 $\upmu$m. This is fully compatible with our expectations. 

Last, equation (\ref{eq:disorderV3b}) predicts that when going from small to large values of $L$ (with respect to $\xi$), $V_\mathrm{3}$ must increase by 0.6V ($\delta V_\mathrm{3}/|V_\mathrm{3}|\approx 0.2$) which is indeed what is observed experimentally (see property P5). 

Overall, the above analysis is fully consistent with smooth density variations being the current bottleneck in our quantitative predictions of pinch-off voltages. To go beyond this limitation, one needs to incorporate information about the {\it local} electronic density within the model. An example of such a procedure is shown in Fig. \ref{fig:map_2D} where we use the experimental value of $\rm V_3$ to calibrate a {\it single parameter} model with $n_\mathrm{s} = n_\mathrm{g}$ (see the discussion of section \ref{sec:fermi}). Figure \ref{fig:map_2D} shows the pinch-off ``phase diagram" as a function of the bottom $V_\mathrm{b}$ and top $V_\mathrm{t}$ gate voltages when these two gates are biased independently. We find that the case $V_\mathrm{b} \ne V_\mathrm{t}$ is quantitatively predicted with an accuracy better than 1\%, \textit{i.e.} significantly improved with respect to a global calibration.

\section{Conclusion}

This article presents a step in the direction of building a precise modeling stack for quantum nanoelectronics. 
Eventually, one aims at high predictive power, so that the simulation tools could be used at the design level of the experiments. A large data set of 110 different quantum point contacts with 48 different designs has been measured. 
This data set has allowed us to perform robust comparisons with the simulations, evaluate the current limitations and improve the modeling and its calibration protocol. At the moment, we have only exploited the pinch-off voltages of the current-gate voltage characteristics. However, the full experimental data set is published together with this article, so that more analysis and more refined simulations may be done later. In particular, it would be interesting to analyse how the conductance plateaus depend on the QPC geometry.

Focusing on the electrostatics, \textit{i.e.} at reconstructing the charge distribution inside the device, we achieved a prediction of the pinch-off voltages with a 5-10\% accuracy when using a single global calibration of the modeling. Multiple aspects in the experiments point to disorder --- slow spatial variations of the electronic density of $\pm  5-10\%$ --- to be the limiting factor of our accuracy. This variation is probably due the presence of inhomogeneities in the dopant layers that leads to spatial variation on a $\mu m$ scale. 

To move forward and obtain more predictive simulations, several strategies can be used. The simplest is to design samples small enough ($\le$ 2 $\upmu$m) so that the calibration of the model can be done in situ using \textit{e.g.} the value of $V_3$. For larger samples, one may design them so that the density in different parts of the device may be calibrated separately using independent gates. An even more advanced approach would use machine learning in order to account for the disorder. Steps in this direction have already been taken in Ref. \cite{Percebois2021} where the authors used deep learning and scanning gate microscopy images to reconstruct an underlying disorder potential. 

Being in possession of a reliable model for the electrostatics of the system opens the possibility for accurate quantum transport simulations. They could then be used for optimizing various figures of merit at the design level of the experiment. Such approaches will become increasingly important in quantum nanoelectronics, in particular as one scales up to increasingly more complex devices \cite{Roussely, Takada} .

\section*{Acknowledgment}
Warm thanks to Ulf Genser, Anne Anthore, Fr\'ederic Pierre,  Arne Ludwig and Andreas D. Wieck for in-depth discussions about dopants and charge distribution in GaAs/AlGaAs heterostructures. This project has been funded by the European Union's Horizon 2020 research and innovation programme under grant agreement No 862683 (UltraFastNano).
E.C. acknowledges the European Union's Horizon 2020 research and innovation program under the Marie Skłodowska-Curie grant agreement No. 840550. J.W. acknowledges the European Union's Horizon 2020 research and innovation program under the Marie Skłodowska-Curie grant agreement No 754303. C.B. and X.W. acknowledge financial support from joint JST - ANR Research Project through Japan Science and Technology Agency, CREST (grant number JPMJCR1876) and the French Agence Nationale de la Recherche, Project QCONTROL ANR-18-JSTQ-0001.

\appendix

\section{Tables of extracted pinch-off}
\label{sec:suppl}

In this appendix, we collect the values of the different pinch-off voltages $V_\mathrm{1}$, $V_\mathrm{2}$ and $V_\mathrm{3}$ that have been extracted from the experimental $\rm I(V_g)$ curves. The raw experimental data can be found in \cite{dataset}.

\begin{widetext}

\begin{table}[b!]
\begin{center}
\begin{tabular}{ | c | c | c | c | c | c | c | c | c | c | c | c | }
\hline
\rowcolor{lightgray} QPC & W(nm) & L(nm) & V3a & V2a & V1a & V3b & V2b & V1b & V3c & V2c & V1c\\
A1 & 250 & 50 & -2.10 &	-0.86 &	-0.44 &	-1.95 & -0.86 &	-0.45 &	-2.20 &	-0.89 &	-0.46 \\
A2 & 300 & 100 & -2.09 & -0.88 & -0.44 & -1.87 & -0.88 & -0.45 & -2.02 & -0.90 & -0.46 \\
A3 & 300 & 250	& -1.41 & -0.88 & -0.45 & -1.29 & -0.87 & -0.45 & -1.52 & -0.89 &  -0.46 \\
A4 & 300 & 500 & -1.22 & -0.89 & -0.45 & -1.17 & -0.87 & -0.45 & -1.22 & -0.89 & -0.46 \\
A5 & 500 & 1000 & -1.96 & -0.88 & -0.45 & -1.84 & -0.88 & -0.45 & -2.05 & -0.90 & -0.46 \\
A6 & 500 & 2500 & -1.46 & -0.86 & -0.45 & -1.82 & -0.86 & -0.45 & -1.97 & -0.91 & -0.46 \\
A7 & 500 & 5000 & -1.83 & -0.88 & -0.44 & - & - & - & - & - & - \\	
A8 & 500 & 1e4 & -1.79 & -0.88 & -0.45 & - & - & - & - & - & - \\
\hline
\end{tabular}
\end{center}
\end{table}

\begin{table}[b!]
\begin{center}
\begin{tabular}{ | c | c | c | c | c | c | }
\hline
\rowcolor{lightgray} QPC & W(nm) & L(nm) & V3d & V2d & V1d \\
A1 & 250 & 50 & -2.00 & -0.93 & -0.47 \\
A2 & 300 & 100 & -2.13 & -0.95 & -0.47 \\
A3 & 300 & 250 & -1.47 & -0.94 & -0.47 \\
A4 & 300 & 500 & -1.24 & -0.93 & -0.46 \\
A5 & 500 & 1000 & -1.97 & -0.94 & -0.46 \\
A6 & 500 & 2500 & -1.90 & -0.90 & -0.46 \\
\hline
\end{tabular}
\end{center}
\end{table}

\begin{table}[b!]
\begin{center}
\begin{tabular}{| c | c | c | c | c | c | c | c | c | c | c | c |}
\hline
\rowcolor{lightgray} QPC & W(nm) & R(nm) & V3a & V2a & V1a & V3b & V2b & V1b & V3c & V2c & V1c \\
B1 & 250 & 25 & -1.94 & -0.88 & -0.43 & -2.43 & -0.95 & -0.48 & - & - & - \\
B2 & 300 & 50 & -2.35 & -0.88 & -0.44 & -2.42 & -0.95 & -0.48 & - & - & - \\		
B3 & 300 & 125 & -1.71 & -0.88 & -0.44 & -1.77 & -0.96 & -0.48 & -1.59 & -0.94 & -0.47 \\
B4 & 300 & 250 & -1.51 & -0.86 & -0.44 & -1.57 & -0.96 & -0.49 & - & - & - \\ 
B5 & 500 & 500 & -2.31 & -0.88 & -0.44 & -2.49 & -0.98 & -0.50 & - & - & -  \\	
B6 & 500 & 1250 & -1.98 & -0.87 & -0.44 & -2.20 & -0.97 & -0.49 & - & - & - \\		
B7 & 500 & 2500 & -1.97 & -0.86 & -0.45 & -2.08 & -0.97 & -0.50 & -2.00 & -0.91 & -0.47\\ 
B8 & 500 & 5000 & -1.93 & -0.87 & -0.44 & -2.00 & -0.98 & -0.50 & - & - & - \\
\hline
\end{tabular}
\end{center}
\end{table}

\begin{table}[!]
\begin{center}
\begin{tabular}{| c | c | c | c | c | c | c | c | c | c | c | c | c | }
\hline
\rowcolor{lightgray} QPC & W(nm) & R(nm) & L(nm) & V3a & V2a & V1a & V3b & V2b & V1b & V3c & V2c & V1c \\
C1 & 250 & 1000 & 50 & -1.0 & -0.98 & -0.49 & -0.96 & -0.87 & -0.44 & -1.05 & -0.92 & -0.46 \\
C2 & 300 & 1000	& 100 & -1.64 & -0.90 & -0.48 & -1.09 & -0.87 & -0.45 & - & - & - \\			
C3 & 300 & 1000 & 250 & -1.21 & -0.98 & -0.49 & -1.09 & -0.88 & -0.45 & -1.21 &	-0.94 &	-0.46 \\
C4 & 300 & 1000 & 500 & -1.18 & -0.97 &	-0.48 &	-1.08 &	-0.86 &	-0.45 &	-1.15 &	-0.92 & -0.46 \\
C5 & 500 & 1000 & 1000 & -1.98 & -0.93 & -0.48 &      &       &       & -2.02 & -0.89 & -0.46 \\
C6 & 500 & 1000 & 2500 & -1.94 & -0.95 & -0.48 & -1.79 & -0.89 & -0.45 & -1.93 & -0.92 & -0.47 \\
C7 & 500 & 1000 & 5000 & -1.91 & -0.95 & -0.48 & -1.68 & -0.88 & -0.45 & -1.92 & -0.91 & -0.48 \\
C8 & 500 & 1000 & 10000 & -1.85 & -0.94 & -0.48 & -1.70 & -0.88 & -0.45 & -1.87 & -0.92 & -0.48 \\
\hline
\end{tabular}
\end{center}
\caption{\label{tab:V123_exp_small} Experimental pinch-off voltages for the short designs.}
\end{table}

\begin{table}[!]
\begin{center}
\begin{tabular}{| c | c | c | c | c | c | c | c | c | c | c | c | c | }
\hline
\rowcolor{lightgray} QPC & W(nm) & R(nm) & L(nm) & V3a & V2a & V1a & V3b & V2b & V1b \\
A9 & 750 & 0 & 1000 & -3.37 & -0.90 & -0.45 & -3.34 & -0.87 & -0.44 \\
A10 & 750 & 0 & 2500 & -3.17 & -0.89 & -0.45 & -3.10 &	-0.88 & -0.44 \\
A11 & 750 & 0 & 5000 &	-3.03 & -0.90 & -0.45 & -3.00 &	-0.88 &	-0.44 \\
A12 & 750 & 0 & 10000 & -2.99 & \cellcolor{blue!80!yellow!50}{\color[HTML]{FFFFFF} -0.78} & -0.44 & -2.92 & \cellcolor{blue!80!yellow!50}{\color[HTML]{FFFFFF} -0.78} & -0.43 \\
A13 & 750 & 0 & 25000 & -2.89 & \cellcolor{blue!80!yellow!50}{\color[HTML]{FFFFFF} -0.78} & -0.44 & -2.85 & \cellcolor{blue!80!yellow!50}{\color[HTML]{FFFFFF} -0.77} & -0.43 \\
A14 & 750 & 0 & 50000 & -1.96 & \cellcolor{blue!80!yellow!50}{\color[HTML]{FFFFFF} -0.77} & -0.44 & -2.72 & \cellcolor{blue!80!yellow!50}{\color[HTML]{FFFFFF} -0.78} & -0.43 \\
A15 & 1000 & 0 & 10000 & -4.1 & -0.90 & -0.44 & -4.18 & -0.91 & -0.44 \\
A16 & 1000 & 0 & 50000 & -3.92 & \cellcolor{blue!80!yellow!50}{\color[HTML]{FFFFFF} -0.78} & -0.42 & - & \cellcolor{blue!80!yellow!50}{\color[HTML]{FFFFFF} -} & - \\			
\hline
B9 & 750 & 500 & 0 & -3.81 & -0.86 & -0.44 & -4.00 & -0.90 & -0.47 \\		
B10 & 750 & 1250 & 0 & -3.50 & -0.89 & -0.44 & -3.40 & -0.86 & -0.46 \\			
B11 & 750 &	2500 & 0 & -3.25 & -0.89 & -0.44 & -3.18 & -0.88 & -0.45 \\			
B12 & 750 & 5000 & 0 & -3.21 & \cellcolor{blue!80!yellow!50}{\color[HTML]{FFFFFF} -0.77} & -0.43 & -3.10 & \cellcolor{blue!80!yellow!50}{\color[HTML]{FFFFFF} -0.75} & -0.45 \\			
B13 & 750 &	12500 &	0 &	-3.11 &	\cellcolor{blue!80!yellow!50}{\color[HTML]{FFFFFF} -0.78} & -0.44 & -3.02 & \cellcolor{blue!80!yellow!50}{\color[HTML]{FFFFFF} -0.77} & -0.44 \\			
B14 & 750 & 25000 & 0 & -3.08 & \cellcolor{blue!80!yellow!50}{\color[HTML]{FFFFFF} -0.78} & -0.44 & -3.04 & \cellcolor{blue!80!yellow!50}{\color[HTML]{FFFFFF} -0.78} & -0.44 \\			
B15 & 1000 & 5000 &	0 & -4.00 &	-0.90 & -0.45 & -7.0 & -0.90 & -0.46 \\			
B16 & 1000 & 25000 & 0 & -3.47 & \cellcolor{blue!80!yellow!50}{\color[HTML]{FFFFFF} -0.80} & -0.44 & - & \cellcolor{blue!80!yellow!50}{\color[HTML]{FFFFFF} -} & - \\
\hline
C9 & 750 & 1000 & 1000 & -3.07 & -0.87 & -0.45 & -3.20 & -0.89 & -0.45 \\
C10 & 750 & 1000 & 2500 & -2.96 & -0.90 & -0.45 & -3.02 & -0.89 & -0.45 \\
C11 & 750 & 1000 & 5000 & -2.92 & -0.89 & -0.45 & -1.10 & -0.89 & -0.45 \\
C12 & 750 & 1000 & 10000 & -1.73 & \cellcolor{blue!80!yellow!50}{\color[HTML]{FFFFFF} -0.77} & -0.44 & -2.93 & \cellcolor{blue!80!yellow!50}{\color[HTML]{FFFFFF} -0.77} & -0.45 \\
C13 & 750 & 1000 & 25000 & -0.78 & \cellcolor{blue!80!yellow!50}{\color[HTML]{FFFFFF} -0.76} & -0.44 & -2.82 & \cellcolor{blue!80!yellow!50}{\color[HTML]{FFFFFF} -0.77} & -0.45 \\
C14 & 750 & 1000 & 50000 & -1.91 & \cellcolor{blue!80!yellow!50}{\color[HTML]{FFFFFF} -0.76} & -0.43 & - & \cellcolor{blue!80!yellow!50}{\color[HTML]{FFFFFF} -} & - \\	
C15 & 1000 & 1000 & 10000 & -4.13 & -0.91 & -0.46 & -4.17 & -0.93 & -0.47 \\
C16 & 1000 & 1000 & 50000 &	-3.88 & \cellcolor{blue!80!yellow!50}{\color[HTML]{FFFFFF} -0.79} & -0.45 & -3.55 & \cellcolor{blue!80!yellow!50}{\color[HTML]{FFFFFF} -0.78} & -0.44 \\
\hline
\end{tabular}
\caption{\label{tab:V123_exp_large} Experimental pinch-off voltages for the long designs. Designs that have $W_\mathrm{g} = 80$ nm in the narrow gate region are highlighted in blue. The rest have $W_\mathrm{g} = 50$ nm like the short designs.}
\end{center}
\end{table}

\begin{figure*}
    \centering
    \includegraphics[scale=0.55]{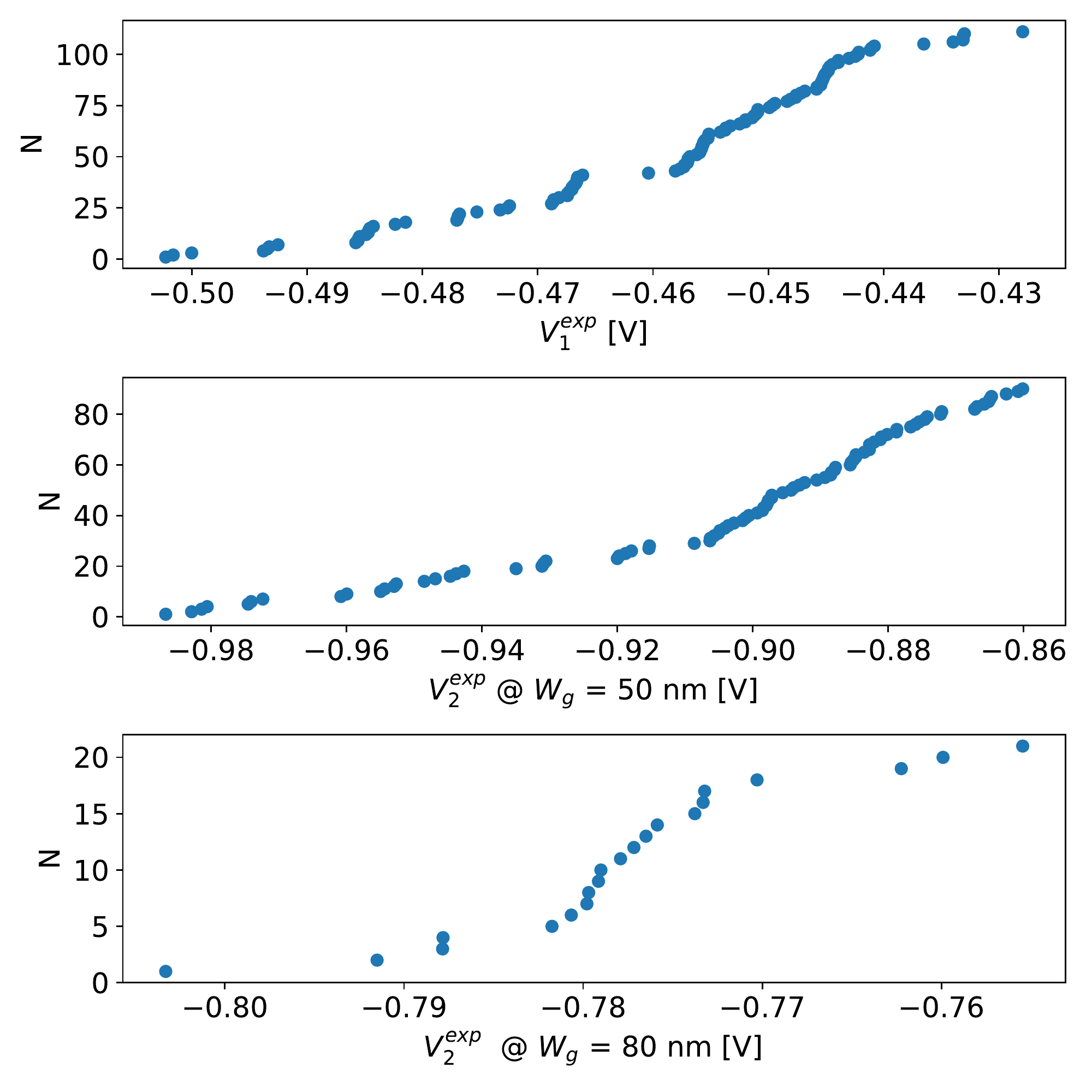}
    \caption{   }
    \label{fig:Vp_CDF} Cumulative distribution of $V_\mathrm{1}$ (top), $V_\mathrm{2}$ with $W_\mathrm{g}=50$ nm (middle) and $V_\mathrm{2}$ with $W_\mathrm{g} = 80$ nm (bottom) for the entire set of measured samples. The N versus x plots show the number of samples N(x) whose pinch-off voltage $V_\mathrm{1}$/$V_\mathrm{2}$ is smaller than x. 
\end{figure*}

\end{widetext}

\clearpage
\newpage

\bibliographystyle{unsrt}
\bibliography{main}

\end{document}